\documentclass[%
 aip,
rsi,%
 amsmath,amssymb,
preprint,%
]{revtex4-1}

\usepackage{diagbox}
\usepackage{graphicx}
\usepackage{dcolumn}
\usepackage{bm}
\usepackage[usenames,dvipsnames]{color}

\usepackage{enumitem}
\usepackage{tabto}
\begin{document}

\global\long\def\R{\mathbb{R}}

\global\long\def\Rn{\mathbb{R}^{n}}
\global\long\def\Zn{\mathbb{Z}^{n}}

\global\long\def\eps{\varepsilon}

\global\long\def\ue{u^{\eps}}

\global\long\def\bQ{\boldsymbol{Q}}
\global\long\def\cS{\mathcal{S}}

\global\long\def\cL{\mathcal{F}}

\global\long\def\cF{\mathcal{L}}
\global\long\def\d{\mathrm{d}}

\preprint{AIP/123-QED}

\title{Estimation of the infinitesimal generator by square-root approximation}

\author{Luca Donati}
 \email{luca.donati@fu-berlin.de} 
 \affiliation{Department of Biology, Chemistry, Pharmacy, Freie Universit\"at Berlin, Takustra\ss e 3, D-14195 Berlin, Germany}
\author{Martin Heida}
 \email{martin.heida@wias-berlin.de} 
 \affiliation{Weierstrass Institute for Applied Analysis and Stochastics, Mohrenstr. 39, 10117 Berlin, Germany}
\author{Bettina G. Keller}
 \email{bettina.keller@fu-berlin.de} 
 \affiliation{Department of Biology, Chemistry, Pharmacy, Freie Universit\"at Berlin, Takustra\ss e 3, D-14195 Berlin, Germany}
\author{Marcus Weber}%
 \email{weber@zib.de}
\affiliation{Zuse Institute Berlin, Takustr. 7, 14195 Berlin, Germany
}%

\date{\today}

\begin{abstract}
For the analysis of molecular processes, the estimation of time-scales, i.e., transition rates, is very important.
Estimating the transition rates between molecular conformations is -- from a mathematical point of view -- an invariant subspace projection problem. A certain infinitesimal generator acting on function space is projected to a low-dimensional rate matrix. 
This projection can be performed in two steps. First, the infinitesimal generator is discretized, then the invariant subspace is approximated and used for the subspace projection. 
In our approach, the discretization will be based on a Voronoi tessellation of the conformational space. We will show that the discretized infinitesimal generator can simply be approximated by the geometric average of the Boltzmann weights of the Voronoi cells.
Thus, there is a direct correlation between the potential energy surface of molecular structures and the transition rates of conformational changes.
We present results for a 2d-diffusion process and Alanine dipeptide.

\end{abstract}

\maketitle

\section{Introduction}

Molecular processes, like conformational changes or binding processes, have been modeled as time-continuous processes switching between a finite number of molecular states \cite{pande10}.  In continuous space, the transition behavior can be described by a propagator\cite{Schuette1999b}. By Galerkin's method, this propagator is projected to a finite dimensional matrix.
A wide collection of tools\cite{Prinz2011} have been developed to numerically estimate the  discretized version of the propagator, i.e., the transition probability matrix which acts on small subsets of the (discretized) conformational space. This transition matrix and the corresponding Markov chain have a unique stationary distribution, as long as the process is ergodic and aperiodic. 
Associated to the propagator, there exists another operator, named infinitesimal generator\cite{Schuette1999b, Oksendal2003}. The infinitesimal generator provides transition pattern in terms of rates in conformational space. The discretized version of the generator is, thus, called rate matrix.
For estimating this rate matrix, we will apply the square root approximation in the following. This approximation has been described in literature before\cite{Lie2013, schild_thesis}. It had been conjectured, that this approximation provides a meaningful rate matrix. Just recently\cite{Heida2017} it has been proved, indeed, that a rate matrix built on that approximation and on an arbitrary tessellation of the space, like a Voronoi tessellation, converges to the backward generator of the Smoluchowski equation, that describes the dynamics of a molecular system in the limit of high friction.
In this paper, we give a detailed explanation of the mentioned operators that describe the dynamics of a molecular system. We recall the main steps to prove that the square root approximation converges to an infinitesimal generator in the limit of small Voronoi cells.
This approximation is valid except for one unknown scalar factor -- the flux of the system. Finally, we discuss an approach for determining this flux. In our case, we use the PCCA+ method to build the rate matrix of the conformations (metastable states) and scale the transition rates properly.
We will present results for a two dimensional system and for Alanine-dipeptide.

\section{Theory}
\begin{figure*}[h!]
  \begin{center}
  \includegraphics[scale=1.0]{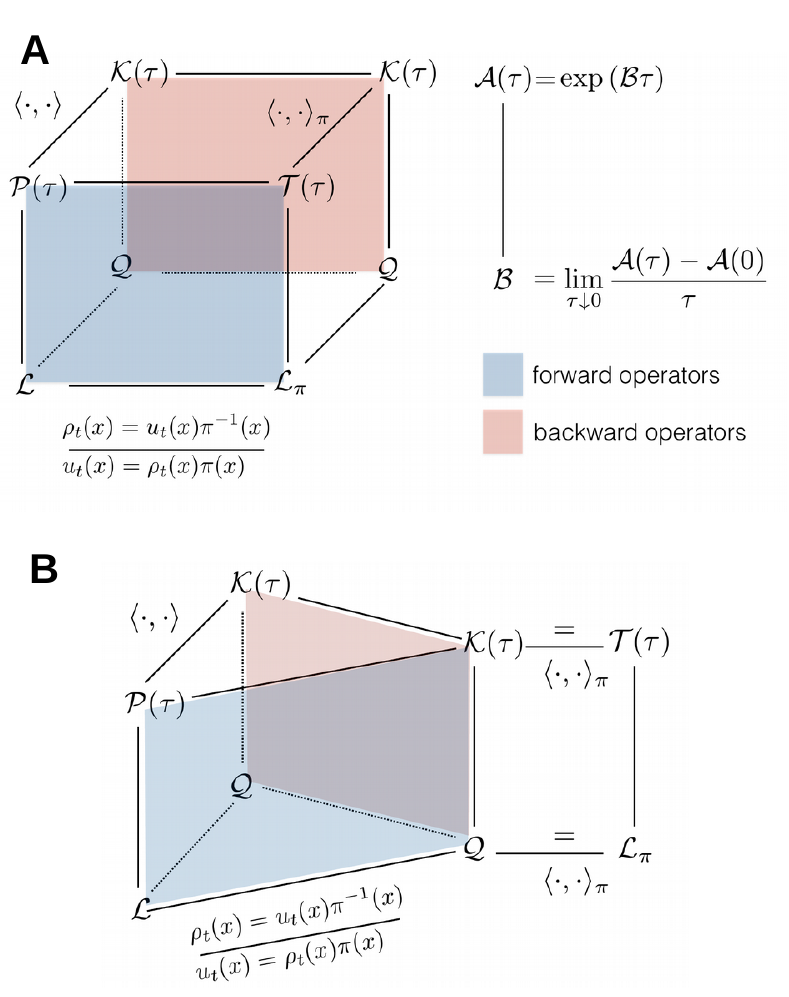}
    \caption{Scheme summarizing the theory of the transfer operator, the generator and the associated transition matrices.}
    \label{fig:op1}
  \end{center}
\end{figure*}

To fully explain the theoretical background, we need to introduce a set of related operators (fig.~\ref{fig:op1}). In section \ref{ii-a}, we discuss the relation between the transfer operators (top face of the blocks in fig.~\ref{fig:op1}-A and ~\ref{fig:op1}.B ), and in section \ref{ii-b}, we discuss the corresponding infinitesimal generators (bottom face of the blocks in fig.~\ref{fig:op1}-A and ~\ref{fig:op1}.B).

\subsection{Propagator,  Koopman operator, and transfer operator}
\label{ii-a}
Let $\lbrace \mathbf{x}_t \rbrace \in \Gamma$ be a dynamic process defined on a state space $\Gamma$.
A large collection of these processes (or equivalently: systems) at time $t$ is called an ensemble, 
and the distribution of the processes across $\Gamma$ at time $t$ is given by the probability density function $\rho_t(\mathbf{x}): \Gamma \rightarrow \mathbb{R}_{\ge 0}$.
Assuming that the process is \emph{Markovian}, we can introduce a continuous operator, called \emph{propagator}\cite{Schuette1999b,Prinz2011} (or \emph{Perron-Frobenius operator}) $\mathcal{P}(\tau):L^1(\Gamma) \rightarrow L^1(\Gamma)$ such that 
\begin{equation}
\label{propagator}
\rho_{t+\tau}(\mathbf{y}) = \mathcal{P}(\tau)  \rho_t(\mathbf{y}) = \int_{\Gamma} p(\mathbf{x},\mathbf{y}; \tau) \rho_t(\mathbf{x}) \mathrm{d} \mathbf{x}
\end{equation}
where $p(\mathbf{x},\mathbf{y}; \tau)$ is the transition probability density, i.e. the conditional probability to find the process in state $\mathbf{y}$ after a \emph{lag time} $\tau$, given the initial state $\mathbf{x}$. 
Thus, the operator $\mathcal{P}(\tau)$ propagates probability densities $\rho_t(\mathbf{x})$ forward in time.

The  \emph{backward transfer operator} \cite{Oksendal2003,Schuette1999b,Huisinga2001} (or \emph{Koopman operator}) $\mathcal{K}(\tau):L^\infty(\Gamma)\rightarrow L^\infty (\Gamma)$ is defined as
\begin{equation}
\label{Koopman_operator}
\mathbb{E}_\mathbf{x} \left[ f(\mathbf{x}_t) \right]
= \mathcal{K}(\tau) f(\mathbf{x}) = \int_{\Gamma} p(\mathbf{x},\mathbf{y}; \tau) f(\mathbf{y})  \mathrm{d} \mathbf{y} 
\end{equation}
where the function $f:\Gamma \rightarrow \mathbb{R}$ is an observable of the system, and $\mathbb{E}_\mathbf{x} \left[ f(\mathbf{x}_t) \right]$ 
is the expected value of this observable at time $t+\tau$, given that the process has been started in $\mathbf{x}_t= \mathbf{x}$ at time $t$.
Thus, the initial point $\mathbf{x}$ is fixed, and the integral in eq. \eqref{Koopman_operator} is computed over the arrival points $\mathbf{y}$.

The Koopman operator $\mathcal{K}(\tau)$ is the left-adjoint of the propagator $\mathcal{P}(\tau)$
\begin{eqnarray}
	\langle \mathcal{K}(\tau) f, \rho_t \rangle 
	 &=& \int_{\Gamma}
	\left[\int_{\Gamma} p(\mathbf{x}, \mathbf{y};\tau) f(\mathbf{y})\, \mathrm{d}\mathbf{y}\right]
	\rho_t(\mathbf{x})\, \mathrm{d}\mathbf{x} \cr
	 &=& \int_{\Gamma}f(\mathbf{y})
	\left[\int_{\Gamma} p(\mathbf{x}, \mathbf{y};\tau)\rho_t(\mathbf{x})\, \mathrm{d}\mathbf{x}  \right]
	\, \mathrm{d}\mathbf{y}\cr \cr
	&=&\langle  f,\mathcal{P}(\tau) \rho_t \rangle \, .
\label{eq:adjointness}	
\end{eqnarray}
In fig.~\ref{fig:op1}, the adjointness is visualized by the edges between the red and the blue face. 
Note that the expression in eq.~\ref{eq:adjointness} can be interpreted as the time evolution of the ensemble average of $f$, 
given that the ensemble was initially distributed according to $\rho_t(x)$ at time $t$
\begin{eqnarray}
	\mathbb{E}[f(\mathbf{x_t})] 
	&=& \int_{\Gamma} f(\mathbf{y}) \rho_{t+\tau}(\mathbf{x}) \, \mathrm{d}\mathbf{x} \cr
	 &=& \int_{\Gamma}f(\mathbf{y})
	\left[\int_{\Gamma} p(\mathbf{x}, \mathbf{y};\tau)\rho_t(\mathbf{x})\, \mathrm{d}\mathbf{x}  \right]
	\, \mathrm{d}\mathbf{y} \cr \cr
	&=& \langle  f,\mathcal{P}(\tau) \rho_t \rangle = \langle \mathcal{K}(\tau) f, \rho_t \rangle \, .
\end{eqnarray}

If the process is \emph{ergodic} and \emph{time-homogeneous} (i.e.~if there are no time-dependent external forces), 
then there exists a unique invariant measure $\pi(\mathbf{x})$ such that for all lag times $\tau$:
\begin{equation}
\pi(\mathbf{y}) =\mathcal{P}(\tau)  \pi(\mathbf{y}) = \int_{\Gamma} p(\mathbf{x},\mathbf{y}; \tau) \pi(\mathbf{x}) \mathrm{d} \mathbf{x} \, .
\end{equation}
For a canonical ensemble (NVT), 
the invariant measure is given by the Boltzmann probability density:
\begin{equation}
\pi(\mathbf{x}) =  \frac{\exp(- \beta \mathcal{H}(\mathbf{x}))}{Z} , \quad  Z = \int_{\Gamma} \exp(- \beta \mathcal{H}(\mathbf{x})) \mathrm{d} \mathbf{x},
\end{equation}
where $\mathcal{H}(\mathbf{x})$ is the classical Hamiltonian of the system, $\beta=\frac{1}{k_B T}$, $T$ is the temperature and $k_B$ the Boltzmann constant.

Using the invariant measure, we define the \emph{forward transfer operator} \cite{Prinz2011}
$\mathcal{T}(\tau) : L_\pi^1(\Gamma) \rightarrow L_\pi^1(\Gamma)$:
\begin{equation}
\label{transfer_operator1}
u_{t+\tau}(\mathbf{y}) 
= \mathcal{T}(\tau)  u_t(\mathbf{y}) 
= \frac{1}{\pi(\mathbf{y})}\int_{\Gamma} p(\mathbf{x},\mathbf{y}; \tau) u_t(\mathbf{x}) \pi(\mathbf{x}) \mathrm{d} \mathbf{x}
\end{equation}
which propagates relative probability densities $u_t(\mathbf{x})$ forward in time, 
where $u_{t+\tau}(\mathbf{y}) = \rho_t (\mathbf{y}) / \pi(\mathbf{y})$. 
Operators which act on probability densities are represented on the left face in fig.~\ref{fig:op1}-A, while operators which act on weighted  probability densities are represented on the right face.
For \emph{reversible} processes, i.e. if
\begin{equation}
\pi(\mathbf{x}) p(\mathbf{x},\mathbf{y}; \tau) = \pi(\mathbf{y}) p(\mathbf{y},\mathbf{x}; \tau) \, ,
\end{equation}
the forward transfer operator and the Koopman operator are self-adjoint with respect to the weighted scalar product 
\begin{equation}
\langle u, v \rangle_{\pi} = \int_\Gamma u(\mathbf{x}) \pi(\mathbf{x}) v(\mathbf{x}) \mathrm{d}\mathbf{x} \, ,
\label{eq:weightedScalarProduct}
\end{equation}
which can be seen by
\begin{eqnarray}
\langle u, \mathcal{T}(\tau) v \rangle_{\pi} 
&=& \int_{\Gamma} u(\mathbf{y}) \pi(\mathbf{y}) \left[ \frac{1}{\pi(\mathbf{y})}\int_{\Gamma} p(\mathbf{x},\mathbf{y}; \tau) \pi(\mathbf{x}) v(\mathbf{x}) \,\mathrm{d}\mathbf{x} \right]\, \mathrm{d}\mathbf{y} \cr
&=& \int_{\Gamma} \int_{\Gamma} u(\mathbf{y}) \pi(\mathbf{x}) p(\mathbf{x},\mathbf{y}; \tau)  v(\mathbf{x}) \,\mathrm{d}\mathbf{x} \, \mathrm{d}\mathbf{y} \cr
&=& \int_{\Gamma} \int_{\Gamma} u(\mathbf{y}) \pi(\mathbf{y}) p(\mathbf{y},\mathbf{x}; \tau)  v(\mathbf{x}) \,\mathrm{d}\mathbf{x} \, \mathrm{d}\mathbf{y} \cr
&=& \int_{\Gamma} \left[\frac{1}{\pi(\mathbf{x})} \int_{\Gamma}   p(\mathbf{y},\mathbf{x}; \tau) \pi(\mathbf{y}) u(\mathbf{y}) \,\mathrm{d}\mathbf{y} \right] \pi(\mathbf{x}) v(\mathbf{x})  \, \mathrm{d}\mathbf{x} \cr \cr
&=& \langle \mathcal{T}(\tau) u, v  \rangle_{\pi} \, .
\label{eq:selfAdjointForwardTransferOperator}
\end{eqnarray}
Note that an analogous calculation holds for Koopman operator $\mathcal{K}(\tau)$. The most important insight is that for reversible processes we find $\mathcal{K}(\tau)=\mathcal{T}(\tau)$ i.e.
\begin{equation}\label{eq:K equals Tfwd}
\frac{1}{\pi(\mathbf{y})}\int_{\Gamma} p(\mathbf{x},\mathbf{y}; \tau) \pi(\mathbf{x}) v(\mathbf{x}) \,\mathrm{d}\mathbf{x} =\int_{\Gamma} p(\mathbf{y},\mathbf{x}; \tau) v(\mathbf{x}) \,\mathrm{d}\mathbf{x}\,.
\end{equation}
Thus, for reversible processes, the right face in fig.~\ref{fig:op1}-A. "collapses" into a single line (fig.~\ref{fig:op1}-B).
This implies that, for reversible processes, the transfer operator is the adjoint of the propagator with respect to the Euclidean scalar product
\begin{eqnarray}
	\langle u, \mathcal{P}(\tau) v \rangle &=& \langle \mathcal{K}(\tau)  u, v\rangle =  \langle \mathcal{T}(\tau)  u, v\rangle \, .
\label{eq:adjointPropagatorTransferOperators}
\end{eqnarray}
The expression in eq.~\ref{eq:adjointPropagatorTransferOperators} can be interpreted as the correlation function 
$\mbox{cor}(u,v; \tau)$ between $u(\mathbf{x})$ and $v(\mathbf{x})$.

%
\subsection{Evolution equation and the generator}
\label{ii-b}
The time-derivative of the probability density at time $t=0$ is then given as
\begin{eqnarray}
 \left. \dot{\rho_t}(\mathbf{x}) \right\vert_{t=0} 
 &=&  	\left. \frac{\partial }{\partial t} \rho_t(\mathbf{x}) \right\vert_{t=0} \cr
 &=&		\lim_{\tau\rightarrow 0}\frac{\rho_{\tau}(\mathbf{x})- \rho_0(\mathbf{x})}{\tau} \cr
 &=& 	\lim_{\tau\rightarrow 0}\frac{\mathcal{P}(\tau)\rho_0(\mathbf{x})- \mathcal{P}(0)\rho_0(\mathbf{x})}{\tau} \cr
 &=& 	\lim_{\tau\rightarrow 0}\frac{\mathcal{P}(\tau)- \mathcal{P}(0)}{\tau}  \rho_0(\mathbf{x}) \cr 
 &=& 	\mathcal{L}\rho_0(\mathbf{x}) \, .
\label{eq:timeEvolutionOperator01} 
\end{eqnarray}
where $\mathcal{L}$ is the infinitesimal generator of $\mathcal{P}(\tau)$. $\mathcal{L}$ is a function of the time-derivative of the transition probability density
\begin{eqnarray}
	\mathcal{L}\rho_0(\mathbf{x}) 
	&=& \int_{\Gamma}\lim_{\tau\rightarrow 0}\frac{p(\mathbf{x}, \mathbf{y}; \tau) - p(\mathbf{x}, \mathbf{y}; 0) }{\tau} \rho_0(\mathbf{x}) \, \mathrm{d}\mathbf{x} \cr
	&=& \int_{\Gamma} \left.\frac{\partial }{\partial \tau}p(\mathbf{x}, \mathbf{y}; \tau)\right\vert_{\tau=0}\rho_0(\mathbf{x}) \, \mathrm{d}\mathbf{x} \cr
	&=& \left.\frac{\partial }{\partial \tau} \mathcal{P}(\tau)\right\vert_{\tau=0}\rho_0(\mathbf{x}) \, ,
\label{eq:timeEvolutionOperator02} 
\end{eqnarray}
and the time evolution equation is
\begin{eqnarray}
\dot{\rho_t}(\mathbf{x}) = \mathcal{L}\rho_t(\mathbf{x}) \, .
\label{eq:evolution equation}	 
\end{eqnarray}
If the dynamic process $\lbrace \mathbf{x}_t \rbrace$ evolves according to the Langevin equation or the Brownian equation of motion, 
eq.~\ref{eq:timeEvolutionOperator02} represents the Fokker-Planck equation. 
If the process evolves according to Hamiltonian dynamics, $\mathcal{L}$ represents the classical Liouville operator.
The relation between an operator and its infinitesimal generator is represented by vertical lines in fig.~\ref{fig:op1}.
The left-adjoint of $\mathcal{L}$ is $\mathcal{Q}$, the \emph{infinitesimal generator} $\mathcal{K}(\tau)$
\begin{eqnarray}
	\langle f, \mathcal{L} \rho_0 \rangle
	&=&  \int_{\Gamma} f(y)
		\left[\int_{\Gamma} \left.\frac{\partial }{\partial \tau}p(\mathbf{x}, \mathbf{y}; \tau)\right\vert_{\tau=0}\rho_0(\mathbf{x}) \, \mathrm{d}\mathbf{x}\right]
	 	\, \mathrm{d}\mathbf{y} \cr
	&=&  \int_{\Gamma} 
		\left[\int_{\Gamma} \left.\frac{\partial }{\partial \tau}p(\mathbf{x}, \mathbf{y}; \tau)\right\vert_{\tau=0}f(y) \, \mathrm{d}\mathbf{y}\right]
		\rho_0(\mathbf{x})
	 	\, \mathrm{d}\mathbf{x} \cr 
	&=& 	\langle \mathcal{Q}f, \rho_0 \rangle	
\end{eqnarray}
Analogous to the relation between $\mathcal{L}$ and the propagator $\mathcal{P}(\tau)$ (eq.~\ref{eq:timeEvolutionOperator01} and eq.~\ref{eq:timeEvolutionOperator02}), 
the infinitesimal generator $\mathcal{Q}$ is linked to the Koopman operator $\mathcal{K}(\tau)$ by
\begin{eqnarray}
\mathcal{Q} \, f(\mathbf{x})
& = & \left. \frac{\partial \mathcal{K}(\tau)}{\partial \tau} \right\vert_{\tau=0} f(\mathbf{x}) \\
& = & \lim_{\tau \downarrow 0} \frac{\mathcal{K}(\tau) f(\mathbf{x}) - \mathcal{K}(0)f(\mathbf{x}) }{\tau} \\
& = & \lim_{\tau \downarrow 0} \frac{\mathbb{E}_\mathbf{x} \left[ f(\mathbf{x}_t)\right] - f(\mathbf{x}) }{\tau} \\
\label{eq:generator01}
\end{eqnarray}
Thus, we can derive $\mathcal{Q}$ by defining the adjoint of $\mathcal{L}$, or by regarding the time-derivative of $\mathbb{E}_\mathbf{x}\left[f(\mathbf{x}_t)\right]$.
Note that this definition of the infinitesimal generator is consistent with the usual definition of the generator of a diffusion process as
 \cite{Schuette1999b,Huisinga2001,Weber2011,Kube2005, Oksendal2003}.

In an analogous way, we can derive the infinitesimal generator $\mathcal{L}_\pi$ of $\mathcal{T}(\tau)$ by examining the time-derivative of $u_t(\mathbf{x})$. However, for reversible processes, the self-adjointness of $\mathcal{T}(\tau)$ and $\mathcal{K}(\tau)$ with respect to eq. \eqref{propagator}, extends to their infinitesimal generators (fig.~\ref{fig:op1}-B)
\begin{eqnarray}
\mathcal{Q} \, f(\mathbf{x})
& = & \left. \frac{\partial \mathcal{T}(\tau)}{\partial \tau} \right\vert_{\tau=0} f(\mathbf{x}) \\
& = & \lim_{\tau \downarrow 0} \frac{\mathcal{T}(\tau) f(\mathbf{x}) - \mathcal{T}(0)f(\mathbf{x}) }{\tau} \\
& = & \mathcal{L}_\pi f(\mathbf{x})
\label{eq:generator02}
\end{eqnarray}
Note, the term infinitesimal generator $\mathcal{L}$ means that the semigroup of $\mathcal{P}(\tau)$ with $\tau>0$ can be generated by
\begin{eqnarray}
\mathcal{P}(\tau) = \exp (\mathcal{L \tau})
\end{eqnarray}
Analogous relation holds for all other operators/infinitesimal generator pairs in this manuscript.


\subsection{Discretization of the time evolution operator and the generator}
Consider an arbitrary partition of the state space in Voronoi cells
$\Gamma = \cup_{i=1}^n \Omega_i$ with associated indicator functions given by 
\begin{equation} 
\mathbf{1}_i (\mathbf{x}) := \begin{cases} 1, & \mathbf{x} \in \Omega_i \\ 0, & \mathbf{x} \not \in \Omega_i \end{cases}
\end{equation}
Assume that $\rho=\sum_i\rho_i \mathbf{1}_i$ is a piecewise constant function solving \eqref{eq:evolution equation} and for given  $\varphi\in\mathbb{R}^n$ multiply \eqref{eq:evolution equation} with $\varphi:=\sum_i \varphi_i\pi \mathbf{1}_i$. We obtain 
\begin{align}
\sum_i\langle\dot \rho_i \mathbf{1}_i,\,\varphi_i \pi \mathbf{1}_i\rangle 
		&= \langle \mathcal{Q}^\ast \rho,\,\sum_i \varphi_i\mathbf{1}_i\rangle_\pi\nonumber\\
        &= \langle \rho,\,\sum_i \varphi_i\mathcal{Q}\mathbf{1}_i\rangle_\pi\nonumber\\
        &= \sum_i \sum_j \varphi_i \rho_j \langle \mathbf{1}_j,\, \mathcal{Q}\mathbf{1}_i \rangle_\pi\,.
        \label{eq:Q_calc1}
\end{align}
Hence, we obtain 
\begin{equation}
\forall \varphi\in\mathbb{R}^n\quad\sum_i \dot \rho_i \varphi_i = \sum_i \sum_j \varphi_i \rho_j Q_{ij}\,\quad\Leftrightarrow\quad \dot \rho=\rho^T Q\,,
        \label{eq:Q_calc2}
\end{equation}
where 
\begin{equation} 
Q_{ij} = \frac{\langle \mathbf{1}_i , \, \mathcal{Q}\, \mathbf{1}_j  \rangle_{\pi}}{\langle \mathbf{1}_i, \, \mathbf{1}_i \rangle_{\pi}} \, .
\end{equation}

By Galerkin discretization, we can construct also a discretization matrix $\mathbf{T}(\tau)$ of the operator $\cal{T}$ via
\begin{equation} 
T_{ij}(\tau) = \frac{\langle \mathcal{T}(\tau) \, \mathbf{1}_i , \, \mathbf{1}_j  \rangle_\pi}{\langle \mathbf{1}_i, \, \mathbf{1}_i \rangle_\pi} = \frac{\langle \mathbf{1}_i , \, \mathcal{T}(\tau) \, \mathbf{1}_j \rangle_\pi}{\langle \mathbf{1}_i, \, \mathbf{1}_i \rangle_\pi} \, .
\label{galerkin_troperator}
\end{equation}
The entries $T_{ij}(\tau)$ describe the transition probability from a cell $\Omega_i$ to $\Omega_j$ after a time $\tau$:
\begin{equation}
T_{ij}(\tau) = \mathbb{P}[\mathbf{x}(t+\tau) \in \Omega_j \, | \, \mathbf{x}(t) \in \Omega_i]
\label{eq:transitionProb}
\end{equation}
Several methods have been developed to estimate $\mathbf{T}(\tau)$, e.g. by Markov State Models\cite{Prinz2011}.

The matrix $\mathbf{Q}$ is a valid discretization of the operators  $\mathcal{Q}$ and $\mathcal{L}=\mathcal{Q}^\ast$, while the matrix $\mathbf{T(\tau)}$ is a discretization of the operators $\mathcal{P}(\tau)$ and $\mathcal{T}(\tau)=\mathcal{P}(\tau)^\ast$. As one can see from \eqref{eq:Q_calc1} and \eqref{eq:Q_calc2}, when applied from the left, the matrices $\mathbf{Q}$ and $\mathbf{T}(\tau)$ act on the weighted space, when applied from the right, act on the euclidean space.
%

The elements $Q_{ij}$, are the time-derivative of the 
transition probabilities $T_{ij}(\tau)$ 
(eq.~\ref{galerkin_troperator} and \ref{eq:transitionProb})
evaluated at $\tau=0$
\begin{equation}
Q_{ij} =  
\frac{\langle \mathbf{1}_i  , \, \frac{\partial \mathcal{T}(\tau)}{\partial \tau} \, \mathbf{1}_j  \rangle_{\pi}}{\langle \mathbf{1}_i, \, \mathbf{1}_i\rangle_{\pi}}
= \left.\frac{\partial }{\partial \tau}\frac{\langle \mathbf{1}_i , \, \mathcal{T}(\tau) \, \mathbf{1}_j  \rangle_{\pi}}{\langle \mathbf{1}_i, \, \mathbf{1}_i \rangle_{\pi}}\right\vert_{\tau=0}
=
\left.\frac{\partial T_{ij}(\tau)}{\partial \tau} \right\vert_{\tau=0}
\end{equation}
Intuitively, the elements of the diagonal $Q_{ii}$ describe the transition rate out the states $i$; while the elements  $Q_{ij}/Q_{ii}$ describe the transition rate from the cell $\Omega_i$ to $\Omega_j$. 
For infinitesimal small values of the lag time $\tau$, the processes which start in cell $\Omega_i$ at time $t$, can only reach directly neighboring (adjacent) cells by time $t+\tau$. Thus for infinitesimal small $\tau$,  $T_{ij}(\tau)$ and $Q_{ij}$
are zero if $\Omega_i$ and $\Omega_j$ are non-adjacent cells. 
%

%
\subsection{Gauss theorem}
We will use the Gauss theorem as stated in Ref.~\onlinecite{Weber2011} to derive a formulation of the rate matrix $\mathbf{Q}$ that can be estimated numerically. 
\paragraph*{Gauss theorem.} Given a Voronoi tessellation of the state space $\Gamma = \cup_{i=1}^n \Omega_i$ and the discretized transfer operator $\mathbf{T}(\tau)$, the matrix $\mathbf{Q}:=\left.\frac{\partial \mathbf{T}(\tau)}{\partial \tau}\right\vert_{\tau=0} $ satisfies
\begin{equation}
Q_{ij} = \frac{1}{\pi_i} \oint_{\partial\Omega_i \partial\Omega_j} \Phi(\mathbf{z}) \,  \pi(\mathbf{z}) \mathrm{d} S(\mathbf{z}) 
\label{gauss}
\end{equation}
where $\partial \Omega_i \partial \Omega_j$ is the common surface between the cell $\Omega_i$ and $\Omega_j$, $\pi_i$ is the Boltzmann density of the cell $i$ and $\Phi(\mathbf{z})$ denotes the flux of the configurations $\mathbf{z} \in \partial \Omega_i \partial \Omega_j$, through the infinitesimal surface $\partial\Omega_i \partial\Omega_j$.

\paragraph*{Proof}
The transition probability density $p(\mathbf{x},\mathbf{y};\tau)$ describes the probability that the system will visit the state $\mathbf{y}$, starting in $\mathbf{x}$, after a time $\tau$. Because the system will be always in some state, it yields $\int_\Gamma p(\mathbf{x},\mathbf{y};\tau) \mathrm{d}\mathbf{y} = 1$.
Thus the conservation of the probability density can be associated to the mass conservation of a fluid, that moves in the state space $\Gamma$, transporting the  properties of the system.
If we consider an ensemble $S_{\tau,\mathbf{x},n}$ of $n\rightarrow \infty$ trajectories, starting in the state $\mathbf{x}$, of length $\tau$, then we can think to the time evolution of the ensemble, like the time evolution of a continuum (fluid); where each trajectory (each particle of the fluid) will end in a different state $\mathbf{y}$ according to $p(\mathbf{x},\mathbf{y};\tau)$. We write the continuity equation:
\begin{equation}
\left.\frac{\partial p(\mathbf{x},\mathbf{y};\tau)}{\partial \tau}\right\vert_{\tau=0} = - \nabla_{\mathbf{y}} \cdot \mathbf{j}
\end{equation}
where $\mathbf{j} = p(\mathbf{x},\mathbf{y};\tau) \mathbf{v}(\mathbf{x})$ is the density flux  and $\mathbf{v}(\mathbf{x})$ is the flow velocity.
We interpret the flux of the probability density as the probability per unit area per unit time, that a trajectory of the ensemble (particle of the fluid) passes through a surface.
While the flow velocity vector is a vector field that represents the velocity with which the system moves from the state $\mathbf{x}$ to $\mathbf{y}$ (or the velocity of the fluid that moves from $\mathbf{x}$ to $\mathbf{y}$). 
We now use the continuity equation, to rewrite 
$\mathbf{Q}$:
\begin{equation}
\begin{aligned}
\label{gauss_proof.qo}
Q_{ij} & := \left.\frac{\partial T_{ij}(\tau)}{\partial \tau}\right\vert_{\tau=0}  \\
(a) & =   \left.\frac{\partial}{\partial \tau} \left[ \frac{1}{\pi_i}\int_{\Omega_j} \int_{\Omega_i} p(\mathbf{x},\mathbf{y};\tau) \pi(\mathbf{x}) \mathrm{d} \mathbf{x}\mathrm{d} \mathbf{y} \right]\right\vert_{\tau=0}  \\
 & =    \frac{1}{\pi_i}\int_{\Omega_j} \int_{\Omega_i}  \left.\frac{\partial p(\mathbf{x},\mathbf{y};\tau)}{\partial \tau}\right\vert_{\tau=0} \pi(\mathbf{x}) \mathrm{d} \mathbf{x}\mathrm{d} \mathbf{y}  \\
(b) & =    \frac{1}{\pi_i}\int_{\Omega_j} \int_{\Omega_i} - \nabla_{\mathbf{y}} \cdot \mathbf{j}|_{\tau= 0} \ \pi(\mathbf{x}) \mathrm{d} \mathbf{x}\mathrm{d} \mathbf{y}  \\
(c) & =   \frac{1}{\pi_i}\int_{\Omega_j} \oint_{\partial\Omega_i} - \mathbf{j}|_{\tau= 0} \cdot \mathbf{n}_i \  \pi(\mathbf{x}) \mathrm{d} S(\mathbf{x})\mathrm{d} \mathbf{y} \\
(d) & =   \frac{1}{\pi_i}\int_{\Omega_j} \oint_{\partial\Omega_i} - \delta_{\mathbf{x}=\mathbf{y}}\mathbf{v}(\mathbf{x}) \cdot \mathbf{n}_i \  \pi(\mathbf{x}) \mathrm{d} S(\mathbf{x})\mathrm{d} \mathbf{y} \\
(e) & =   \frac{1}{\pi_i} \oint_{\partial\Omega_i \partial\Omega_j} \Phi(\mathbf{z}) \  \pi(\mathbf{z}) \mathrm{d} S(\mathbf{z})  
\end{aligned}
\end{equation}
where we have used:
\begin{enumerate}[label=(\alph*)]
\item The result of the Galerkin discretization of the transfer operator in eq. \eqref{galerkin_troperator}.
\item The continuity equation.
\item The divergence theorem. The vector $\mathbf{n}_i$ is the unit vector normal to the surface $\partial \Omega_i$. 
\item Because $\tau = 0$, we have
\begin{equation}
p(\mathbf{x},\mathbf{y};\tau) = \delta_{\mathbf{x}=\mathbf{y}}
\end{equation}
Then the density flux is $\mathbf{j}|_{\tau = 0} = \delta_{\mathbf{x}=\mathbf{y}}\mathbf{v}(\mathbf{x})$.
\item If $\tau = 0$ only instantaneous transitions between neighbor cells have to be taken in account. Thus, the only points that satisfy $\mathbf{x}=\mathbf{y}$, are the points on the intersecting surface $\partial\Omega_i \partial\Omega_j$. We denote those points, in the integral (e), with $\mathbf{z}$. The quantity $\Phi(\mathbf{z})$ denotes the flux of the configurations $\mathbf{z}$ trough the infinitesimal surface $\partial \Omega_i$. Note that  $\Phi(\mathbf{z}) = - \delta_{\mathbf{x}=\mathbf{y}}\mathbf{v}(\mathbf{x}) \cdot \mathbf{n}_i = \delta_{\mathbf{x}=\mathbf{y}}\mathbf{v}(\mathbf{x}) \cdot \mathbf{n}_j$ where $\mathbf{n}_j$ is the unit vector normal to the surface $\partial \Omega_j$.
\end{enumerate}
\hfill$\square$\\
%
\subsection{Square Root Approximation}
\label{sec:sqra}
Following \cite{Lie2013}, we now show how the matrix \eqref{gauss} can be approximated by Square Root Approximation (SqRA).
First of all, we rewrite eq.\ref{gauss}, multiplying and dividing by the quantity
\begin{equation}
s_{ij} = \oint_{\partial\Omega_i \partial\Omega_j} \pi(\mathbf{z}) \, \mathrm{d} S(\mathbf{z})
\end{equation} 
that represents the Boltzmann density of the intersecting surface $\partial \Omega_i \partial \Omega_j$. Thus $Q_{ij}$ becomes
\begin{equation}
\begin{aligned}
\label{sra.qo}
Q_{ij} = &\frac{s_{ij}}{\pi_i} \oint_{\partial\Omega_i \partial\Omega_j} \Phi(\mathbf{z}) \  \frac{\pi(\mathbf{z})}{s_{ij}} \mathrm{d} S(\mathbf{z})\\
= &\frac{s_{ij}}{\pi_i} \langle\Phi \rangle_{ij}
\end{aligned}
\end{equation}
the quantity $\langle\Phi \rangle_{ij}$ represents the mean value of the flux, trough  $\partial \Omega_i \partial \Omega_j$, if the potential energy function was flat ($V(\cdot) = 0$), weighted by the Boltzmann density of the intersecting surface $\partial \Omega_i \partial \Omega_j$.

In what follows,we make the following assumptions
\begin{itemize}
\item There exists a constant $\hat\Phi$ such that for all $i,j$ it holds $\hat\Phi=\langle \Phi \rangle_{ij}$. 
\item The cells are so small such that $\pi$ is almost constant on every $\Omega_i$ and on every interface $\partial\Omega_i\partial\Omega_j$, i.e. $\pi|_{\Omega_i}\approx\pi_i$ and $\pi|_{\partial\Omega_i\partial\Omega_j}\approx\pi_{ij}$.
\end{itemize}

Due to assumption 1, the matrix $\mathbf{Q}$ can be rewritten as
\begin{equation}
Q_{ij}' = \frac{s_{ij}}{\pi_i} \hat \Phi  
\end{equation}
The quantity $s_{ij}$, i.e. the Boltzmann density of the intersecting surface between two neighbor cells $\Omega_i$ and $\Omega_j$, is a surface integral. Due to assumption, $s_{ij}$ can be approximated by the Boltzmann density $\pi_{ij}$ through 
\begin{equation}
\begin{aligned}
s_{ij} = & \oint_{\partial\Omega_i \partial\Omega_j} \pi(\mathbf{z}) \mathrm{d} S(\mathbf{z}) \\
\approx & \,\pi_{ij}
\end{aligned}
\end{equation}
and 
\begin{equation}
Q_{ij}'' = \frac{\pi(\hat{\mathbf{z}})}{\pi_i} \, \hat \Phi  
\label{eq:sqra_final}
\end{equation}
Since all points on the surface $\partial \Omega_i \partial \Omega_j$ are equidistant from the centers of the cells $\Omega_i$ and $\Omega_j$, we estimate $\pi_{ij}$ by the geometric average
\begin{equation}
\pi_{ij} = \sqrt{\pi_i \, \pi_j}\,.
\end{equation}
Thus, the rate matrix is given as
\begin{equation}
Q_{ij}''' = \frac{\sqrt{\pi_i \, \pi_j}}{\pi_i} \, \hat \Phi  = \sqrt{\frac{\pi_j}{\pi_i}} \, \hat  \Phi
\label{eq:rate_matrix_sqra}
\end{equation}
%
%
If we consider a molecular system in the limit of high friction, the dynamics is governed by the Smoluchowski equation and the stationary distribution depends only on the potential energy function $V(\cdot)$. In this case, the Boltzmann density of the midpoint $\hat{\mathbf{z}}$ can be easily estimated as
\begin{equation}
\pi(\hat{\mathbf{z}}) = \sqrt{\pi_i \, \pi_j} = \frac{\exp\lbrace-\beta \frac{1}{2} \left[V(\Omega_i)+V(\Omega_j)\right]\rbrace}{Z}
\end{equation}
The diagonal entries of $\mathbf{Q}$ are $Q_{ii} = - \sum_{j \neq i} Q_{ij}$, thus the sum of the rows is zero $\sum_j Q_{ij}=0$.
The matrix $\mathbf{Q}$ has also the property that the  ordinary differential equation \eqref{eq:Q_calc2} reads
\begin{equation}\label{eq:discr-time-FP}
\left.\frac{\partial \rho_i}{\partial \tau} \right\vert_{\tau=0} = C \sum_{i \sim j} (\rho_j Q_{ji} - \rho_i Q_{ij})
\end{equation}
where $C$ is a normalization constant and the notation $i \sim j$ denote neighboring cells. 
\subsection{Convergence of the rate matrix}
The square root approximation of the infinitesimal generator $\cal Q$ is based on a Voronoi tessellation of the state space. In principle, it is not clear if this kind of approximation has a physically meaningful limit, whenever the number of Voronoi cells tend to infinity. If this limit exists, it is furthermore not clear whether the limit operator is physically reasonable. An arbitrary refinement strategy of increasing the number of Voronoi cells will probably not converge. However, a recent mathematical study \cite{Heida2017} which will we published separately, shows that under suitable assumptions on the Voronoi tessellation, the square root approximation converges towards the backward generator of the Smoluchowski equation, i.e., towards the Langevin dynamics for the limit of high friction.

In what follows, let $\eps$ denote the maximal diameter of the cells
and for given $\eps>0$ let $\left(P_{i}^{\eps}\right)_{i}$ be the
set of points that generate the Voronoi tessellation and let $\Omega_{i}^{\eps}$ be 
the Voronoi cell that corresponds to $P_{i}^{\eps}$. For any continuous
function $u$, we write $\rho_{i}^{\eps}:=\rho(P_{i}^{\eps})$. In particular,
for the function $\pi(x):=\exp\left(-\beta V(x)\right)$ we write
$v_{i}^{\eps}:=\sqrt{\pi_{i}^{\eps}}$ and using \eqref{eq:sqra_final} we  denote the right hand side
of \eqref{eq:discr-time-FP} as 
\begin{equation}
\left(\cF^{\eps}\rho\right)_{i}^{\eps}:=C_{\eps}\sum_{i\sim j}\left(\rho_{j}^{\eps}\frac{v_{i}^{\eps}}{v_{j}^{\eps}}-\rho_{i}^{\eps}\frac{v_{j}^{\eps}}{v_{i}^{\eps}}\right)\,,\label{eq:def:discrete-Fokker-Planck}
\end{equation}
where $\sum_{i\sim j}$ relates to the sum over all neighboring cells
$\Omega_{j}^{\eps}$ of the cell $\Omega_{i}^{\eps}$ and where we
interpret $\left(\cF^{\eps}\rho\right)_{i}^{\eps}$ as a function which
is constant on every $\Omega_{i}^{\eps}$. Written
in a formal way, for the scaling $C_{\eps}=\eps^{-2}$
and a suitable positive definite symmetric matrix $A_{\hom}\in\R^{n\times n}$
it holds for twice continuously differentiable functions $u$ that\cite{Heida2017}
\begin{equation}
C_{\eps}\sum_{i\sim j}\left(\rho_{j}^{\eps}\frac{v_{i}^{\eps}}{v_{j}^{\eps}}-\rho_{i}^{\eps}\frac{v_{j}^{\eps}}{v_{i}^{\eps}}\right)\to\nabla\cdot\left(A_{\hom}\nabla \rho(x)\right)+\beta\nabla\cdot\left(\rho(x)A_{\hom}\nabla V(x)\right)\quad\mbox{as }\eps\to0\label{eq:convergence-FP}
\end{equation}
provided that $P_{i}^{\eps}\to x$ as $\eps\to0$. In case that the
Voronoi tessellations are isotropic, we find $A_{\hom}=a\mathbb{I}$,
where $a>0$ is a constant and $\mathbb{I}$ is the identity. However,
if for some reason the tessellations are systematically anisotropic,
the right hand side of \eqref{eq:convergence-FP} can be brought into
the form of a classical Fokker-Planck operator via a coordinate transform.
Hence, we also call the right hand side of \eqref{eq:convergence-FP}
a Fokker-Planck operator. In what follows, we will roughly explain
how the convergence \eqref{eq:convergence-FP} can be obtained.

There exists a simpler version of \eqref{eq:def:discrete-Fokker-Planck}
known as \emph{the discrete Laplace operator} $\cL u$ having the
form
\[
\left(\cL^{\eps}\rho\right)_{i}^{\eps}:=C_{\eps}\sum_{i\sim j}\left(\rho_{j}^{\eps}-\rho_{i}^{\eps}\right)\,.
\]
Note that the use of $\cF$ and $\cL$ is opposite in\cite{Heida2017}. It turned out that the understanding of the asymptotic behavior of
$\cL^{\eps}$ is essential for the study of the asymptotic behavior
of $\cF^{\eps}$. 

\begin{figure*}[h!]
\begin{center}
\includegraphics[scale=0.7]{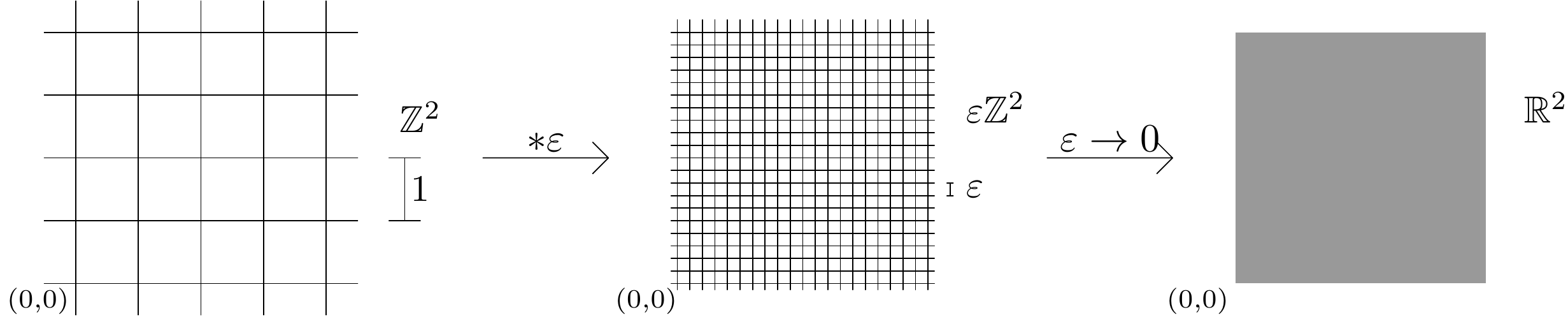}\protect
\caption{The role of the parameter in the case $\protect\eps\mathbb{Z}^{2}$. As $\protect\eps\to 0$, the grid becomes finer and finer and approximates
the whole of $\protect\R^{2}$.
}
\label{fig:eps-Zn}
\end{center}
\end{figure*}

In case the point process is a rectangular grid (Fig. \ref{fig:eps-Zn}),
the operator $\cL^{\eps}$ has been studied intensively and in broad
generality from both physicists (as generator of a markovian process
that models Brownian motion, see the review \cite{Bouchaud1990})
and mathematicians (rigorous results, see the review \cite{Biskup2011review}).
The notion of discrete Laplace operator can be understood as follows:
On the lattice $\eps\Zn$ (that consists of all points $x\in\Rn$
such that $\left(\eps^{-1}x\right)\in\Zn$, see Figure \ref{fig:eps-Zn}),
the \emph{discrete derivative} in the $j$-th direction is given by
$\d_{j,\eps}\rho(x):=\frac{1}{\eps}\left(\rho(x+\eps e_{j})-\rho(x)\right)$,
where $e_{j}$ is the $j$-th unit vector. The \emph{second order
discrete derivative} is given by 
\begin{align}
\d_{j}^{2}\rho(x):&=\frac{1}{\eps}\left(\d_{j}\rho(x)-\d_{j}\rho(x-\eps e_{j})\right)\nonumber\\
	&=\frac{1}{\eps^{2}}\left[\left(\rho(x+\eps e_{j})-\rho(x)\right)+\left(\rho(x-\eps e_{j})-\rho(x)\right)\right]\,.
\end{align}
Hence, we obtain for $x_{i}^{\eps}\in\eps\Zn$ that 
\begin{equation}
\sum_{j=1}^{n}\d_{j}^{2}\rho(x_{i}^{\eps})=\eps^{-2}\sum_{j\sim i}\left(\rho(x_{j}^{\eps})-\rho(x_{i}^{\eps})\right)=\left(\cL^{\eps}\rho\right)_{i}\,,\label{eq:def-discr-laplace-Zn}
\end{equation}
where now $i\sim j$ relates to all neighbors $x_{j}^{\eps}\in\eps\Zn$
of $x_{i}^{\eps}$ s.t. $\left|x_{j}^{\eps}-x_{i}^{\eps}\right|=\eps$.
We now show that for twice continuously differentiable functions
$u$ it holds
\begin{equation}
\left(\cL^{\eps}\rho\right)_{i}^{\eps}\to\Delta \rho\qquad\mbox{as }\eps\to0\mbox{ if }x_{i}^{\eps}\to x\,.\label{eq:conv-discr-laplace-Zn}
\end{equation}
In order to show this, we use Taylor's formula, i.e. 
\begin{equation}
\rho(x+\eps e_{i})-\rho(x)=\partial_{i}\rho(x)\eps+\frac{1}{2}\partial_{i}^{2}\rho(x)\eps^{2}+\sum_{k=3}^{\infty}\frac{1}{k!}\partial_{i}^{k}\rho(x)\eps^{k}
\end{equation}
and hence 
\begin{align}
\d_{j}^{2}\rho(x_{i}^{\eps}) & =\frac{1}{\eps^{2}}\left(\partial_{j}\rho(x_{i}^{\eps})\eps+\frac{1}{2}\partial_{j}^{2}\rho(x_{i}^{\eps})\eps^{2}+\sum_{k=3}^{\infty}\frac{1}{k!}\partial_{j}^{k}\rho(x_{i}^{\eps})\eps^{k}\right)\nonumber \\
 & \quad+\frac{1}{\eps^{2}}\left(-\partial_{j}\rho(x_{i}^{\eps})\eps+\frac{1}{2}\partial_{j}^{2}\rho(x_{i}^{\eps})\eps^{2}+\sum_{k=3}^{\infty}\frac{1}{k!}\partial_{j}^{k}\rho(x_{i}^{\eps})\left(-\eps\right)^{k}\right)\nonumber \\
 & =\partial_{j}^{2}\rho(x_{i}^{\eps})+\mathcal{O}(\eps)\,.\label{eq:discr-snd-deriv}
\end{align}
From this, we obtain that \eqref{eq:conv-discr-laplace-Zn} holds. 

We will now use the above insights to formally understand the asymptotic
behavior of the operator 
\begin{equation}
\left(\cF^{\eps}\rho\right)_{i}:=\eps^{-2}\sum_{i\sim j}\left(\rho_{j}^{\eps}\frac{v_{i}^{\eps}}{v_{j}^{\eps}}-\rho_{i}^{\eps}\frac{v_{j}^{\eps}}{v_{i}^{\eps}}\right)\qquad\mbox{where }v_{i}^{\eps}=\exp\left(-\frac{1}{2}\beta V(x_{i}^{\eps})\right)\label{eq:discr-FP-eps}
\end{equation}
on the lattice $\eps\Zn$. Writing $V_{i}^{\eps}:=V(x_{i}^{\eps})$
and using the Taylor formula we obtain 
\begin{equation}
\frac{v_{i}^{\eps}}{v_{j}^{\eps}}=1-\frac{1}{2}\beta\left(V_{i}^{\eps}-V_{j}^{\eps}\right)+\frac{1}{8}\beta^{2}\left(V_{i}^{\eps}-V_{j}^{\eps}\right)^{2}-\mathcal{O}\left(\left(V_{i}^{\eps}-V_{j}^{\eps}\right)^{3}\right)
\end{equation}
and inserting this expansion into \eqref{eq:discr-FP-eps} we obtain
\begin{align}
\left(\cF^{\eps}\rho\right)_{i} & =\eps^{-2}\sum_{i\sim j}\left(\left(\rho_{j}^{\eps}-\rho_{i}^{\eps}\right)+\frac{\beta}{2}\left(\rho_{j}^{\eps}+\rho_{i}^{\eps}\right)\left(V_{j}^{\eps}-V_{i}^{\eps}\right)\right)\nonumber\\
 & \quad+\eps^{-2}\sum_{i\sim j}\left(\frac{1}{8}\beta^{2}\left(V_{i}^{\eps}-V_{j}^{\eps}\right)^{2}+\mathcal{O}\left(\left(V_{i}^{\eps}-V_{j}^{\eps}\right)^{4}\right)\right)\left(\rho_{j}^{\eps}-\rho_{i}^{\eps}\right)\nonumber\\
 &\quad+\eps^{-2}\sum_{i\sim j}\left(\rho_{j}^{\eps}+\rho_{i}^{\eps}\right)\mathcal{O}\left(\left(V_{i}^{\eps}-V_{j}^{\eps}\right)^{3}\right).
\end{align}
We know that for small $\eps\approx0$ it holds $\left(V_{i}^{\eps}-V_{j}^{\eps}\right)\approx\eps\nabla V(x_{i}^{\eps})+\mathcal{O}(\eps^{2})$
and equivalently $\left(\rho_{i}^{\eps}-\rho_{j}^{\eps}\right)\approx\eps\nabla \rho(x_{i}^{\eps})+\mathcal{O}(\eps^{2})$.
Therefore, we obtain 
\begin{align}
\left(\cF^{\eps}\rho\right)_{i}^{\eps} & =\eps^{-2}\sum_{i\sim j}\left(\left(\rho_{j}^{\eps}-\rho_{i}^{\eps}\right)+\frac{\beta}{2}\left(\rho_{j}^{\eps}+\rho_{i}^{\eps}\right)\left(V_{j}^{\eps}-V_{i}^{\eps}\right)\right)+\mathcal{O}(\eps)\label{eq:discr-FP-eps-expansion-1}
\end{align}
and using once more the Taylor expansion for $u$ and $V$ in \eqref{eq:discr-FP-eps-expansion-1}
we further obtain 
\begin{align}
\left(\cF^{\eps}\rho\right)_{i}^{\eps} & =\eps^{-2}\sum_{i\sim j}\left(\rho_{j}^{\eps}-\rho_{i}^{\eps}\right)+\eps^{-2}\rho_{i}^{\eps}\sum_{i\sim j}\beta\left(V_{j}^{\eps}-V_{i}^{\eps}\right)+\eps^{-2}\sum_{i\sim j}\frac{\beta}{2}\left(\rho_{j}^{\eps}-\rho_{i}^{\eps}\right)\left(V_{j}^{\eps}-V_{i}^{\eps}\right)+\mathcal{O}(\eps)\nonumber \\
 & =\eps^{-2}\sum_{i\sim j}\left(\rho_{j}^{\eps}-\rho_{i}^{\eps}\right)+\eps^{-2}\rho_{i}^{\eps}\sum_{i\sim j}\beta\left(V_{j}^{\eps}-V_{i}^{\eps}\right)+\sum_{i\sim j}\frac{\beta}{2}\partial_{j}\rho(x_{i}^{\eps})\cdot\partial_{j}V(x_{i}^{\eps})+\mathcal{O}(\eps)\,.\label{eq:split-SQRA-Op}
\end{align}

Thus, as $\eps\to0$ we observe that 
\begin{eqnarray*}
\left(\cF^{\eps}\rho\right)_{i}^{\eps} & \to & \Delta \rho(x)+\beta \rho(x)\Delta V(x)+\beta\nabla \rho(x)\cdot\nabla V(x)\qquad x_{i}^{\eps}\to x\\
 & = & \Delta \rho(x)+\beta\nabla\cdot\left(\rho(x)\nabla V(x)\right)
\end{eqnarray*}
on the Grid $\eps\Zn$. Hence, we recover \eqref{eq:convergence-FP}
with $A_{\hom}=1$ for the cubic Voronoi tessellation.

On arbitrary Voronoi tessellations, things become more involved. In
particular, the convergence \eqref{eq:conv-discr-laplace-Zn} or calculations
like in \eqref{eq:discr-snd-deriv} and \eqref{eq:split-SQRA-Op}
do not hold any more, as they rely on the rectangular structure of
$\Zn$. However, the key ideas of the proof remain the same with the
difference that some terms which explicitly cancel out in the above
calculation only vanish in a ``statistically averaged'' sense, using G-convergence. 

G-convergence is a concept from early stage in the
development of Homogenization theory and is extremely rarely used
(refer to \cite{JKO1994}), since other concepts are usually much
better suited. In the discrete setting, G-convergence can be formulated
in the following sense: The operator $\cL^{\eps}$ is called G-convergent
if there exists a symmetric positive definite matrix $A_{\hom}$ such
that for every continuous $f:\,\Omega\to\R$ the sequence $\ue$ of
solutions to 
\begin{equation}
-\left(\cL^{\eps}\rho^\varepsilon\right)_{i}:=-\sum_{j\sim i}\left(\rho_{j}^{\eps}-\rho_{i}^{\eps}\right)=f(P_{i}^{\eps})
\end{equation}
converges in $L^{2}(\Omega)$ to the solutions $\rho$ of $-\nabla\cdot\left(A_{\hom}\nabla \rho\right)=f$,
where we interpret $\rho^\varepsilon$ as a function that is constant on every
cell $\Omega_{i}^{\eps}$. Hence, G-convergence and convergence of
the SQRA-operator are more or less equivalent conditions on the tessellation.
In recent years, G-convergence (or \eqref{eq:conv-discr-laplace-Zn})
has been proved for random operators 
\begin{equation}
\left(\cL_{\omega}^{\eps}\rho\right)(x_{i}):=\frac{1}{\eps^{2}}\sum_{j\sim i}\omega_{ij}\left(\rho(x_{j})-\rho(x_{i})\right)
\end{equation}
on the grid $\eps\Zn$ for a broad range of random coefficients $\omega_{ij}$,
see the overview in \cite{Biskup2011review}. However, for stationary
and ergodic tessellations, the recent results in \cite{Alicandro2011,Heida2017}
seem to be the only ones. 

In conclusion we showed that the convergence \eqref{eq:convergence-FP}
holds on the rectangular grid for sufficiently smooth functions $u$.
The calculations suggest that the result also holds on more general
grids. However, on such more general grids, the mathematics behind
the convergence \eqref{eq:convergence-FP} becomes much more involved
and is thus shifted to the article \cite{Heida2017}. The results
there are though more general as they state that solutions of $\cF^{\eps}\rho^\varepsilon=f^{\eps}$
converge to solutions of $\cF \rho=f$ provided $f^{\eps}\to f$ in $L^{2}(\bQ)$.


\section{Methods}
\subsection{Discretization error}
The Galerkin discretization of the continuous operators $\mathcal{T}(\tau)$ and $\mathcal{Q}$ described in the theory section is one source of errors. 
With regard to the transition probability matrix $\mathbf{T}(\tau)$, the discretization of the space causes the loss of information and the loss of the Markov property \cite{Prinz2011,Weber2011,Kube2008}. 
Let's consider a discretization of the space into $n$ disjoint sets $A_1,\ldots, A_n$, and assume that at time $t>0$, the state $\mathbf{x}(t)$ is in a certain set $A_i$.
The transition probability to the next set at time $t+\tau$ depends on the position  $\mathbf{x}(t)$ inside the set $A_i$.
Although the probability to be in state $\mathbf{x}(t)$ at time $t$ only depends on the previous position $\mathbf{x}(t-\tau)$, this Markov property is lost, on the level of the sets $A_i$. The transition behavior between the sets $A_i$ is not Markovian in general.
The deviation from Markoviantity of the matrix $\mathbf{T}(\tau)$ can be reduced, by choosing a proper partition of the state space. Alternatively, it can be proved that for a large enough value of $\tau$, the implied time scales $t_i = - \frac{\tau}{\log \lambda_i}$ (where $\lambda_i$ are the eigenvalues of the transition probability matrix $\mathbf{T}(\tau)$) converge to constant values as a consequence of the Chapman-Kolmogorov equation.

With regard to the matrix $\mathbf{Q}$, two problems arise: 
\begin{itemize}
\item If $\tau$ is small, the matrix $\mathbf{T}(\tau)$ is not Markovian, and the matrix $\mathbf{Q}$ cannot be considered as a generator of $\mathbf{T}(\tau)$.
\item If $\tau$ is big enough to guarantee the Markovianity of $\mathbf{T}(\tau)$, then the generator $\mathbf{Q}$ is the correct generator, but it is not physically meaningful. A proper generator is defined for $\tau \rightarrow 0$, in other words for instantaneous transitions that occur between neighboring sets. If $\tau$ is too big, then $\mathbf{Q}$ would describe instantaneous transition rates between non-neighboring sets, that are not physical if we are considering a time-continuous dynamics.
\end{itemize}
In conclusion, the matrix $\mathbf{Q}$ is the Galerkin discretization of the infinitesimal generator $\mathcal{Q}$, but it is not the proper generator of the transition probability matrix $\mathbf{T}(\tau)$. 
In the next section, we will see that by using a different basis function, we can discretize the operators in transition matrices that satisfy respectively the Markov property and the generators properties.
\subsection{Transition matrices in the space of conformations}
We now consider a partition of the state-space into $n$ disjoint Voronoi cells and a partition of the configuration space into $n_c$ overlapping conformations, i.e., we introduce a membership matrix $X\in\R^{n\times n_C}$  such that the entry $X_{ij}$ provides the probability that a Voronoi cell $\Omega_i$ belongs to a conformation $C_j$:
\begin{equation}
X_{ij} = X_j(\Omega_i) = \mathbf{P}\left[\Omega_i \in C_j \right] \in [0,1], \quad i=1,\ldots,n; \, j=1,\ldots,n_c
\end{equation} 
The matrix $X$ is nonnegative. Its entries form a partition of the unity, such that the row sum is $\sum_{j=1}^{n_c} X_{ij}=1 \ \ \forall i$. 
In this section, we use the notation $X_{i \bullet}$ to denote row vectors of size $n_c$ and  $X_{\bullet j}$ to denote column vectors of size $n$.
Because of metastability of the conformations, a state $\mathbf{x}$ tends to not leave its starting conformation $C_j$. The product between the rate matrix $\mathbf{Q}$ and the column vector $X_{\bullet j}$ is almost zero for all $j$. 
\begin{equation}
\mathbf{Q} X_{\bullet j} \approx \mathbf{0}
\label{eq:Qchi}
\end{equation}
where $\mathbf{0}$ is the zero vector of size $n$.
For the identification of the conformations it can be assumed, that the column vectors $X_{\bullet j}$ span the same linear space as the leading first $n_c$ right eigenvectors $F_j$ of the matrix $\mathbf{Q}$, with eigenvalues $\theta_j$ near zero \cite{Weber2011}. For writing down the basis transform between these two linear spaces, we call $F$ the matrix $n \times n_c$ of the eigenvectors $F_{\bullet j}$ using the same notation regarding column and row vectors.

We now define an invertible transformation matrix $\mathbf{A}$ of size $n_c \times n_c$, such that $X_{ij} = \sum_{k=1}^{n_c} F_{ik} A_{kj}$ (or $X_{\bullet j} = F A_{\bullet j}$), or in matrix notation $X = F \mathbf{A}$.

Because $\mathbf{Q} F_{\bullet j} = \theta_j F_{\bullet j}$, from eq. \eqref{eq:Qchi}
\begin{equation}
\mathbf{Q} (F \, A_{\bullet j}) = \theta_i (F \, A_{\bullet j}) \approx \mathbf{0}
\end{equation}
Because the rows of the matrix $X$ define the coordinates of $n$ points in $n_c$ dimensions that lie on the standard $n_c$-simplex, we can use Robust Perron Cluster Analysis (PCCA+) \cite{Deuflhard2004} to find the optimal transformation matrix $\mathbf{A}$ and the matrix of the membership functions $X$.
The transition rate matrix $\mathbf{Q}^c$ of the conformations is the Galerkin discretization of the operator $\mathcal{Q}$ on the basis of the membership functions $\chi_j(\mathbf{x})=\sum_{i=1}^n X_{ij}\, {\mathbf{1}}_{\Omega_i}(\mathbf{x})$. Since the membership functions are not orthogonal, the matrix $Q^c$ is the product of two matrices generated by the corresponding inner products: 
\begin{equation}
\label{conf_trans}
\mathbf{Q}^c = \left(\langle \chi, \chi  \rangle_\pi\right)^{-1} \,\langle  \chi, \mathcal{Q} \, \chi  \rangle_\pi 
\end{equation}
%

The following lemma\cite{Kube2006,Weber2009} implies that if $\mathcal{Q}$ is the infinitesimal generator of $\mathcal{T}(\tau)$, then the matrix $\mathbf{Q}^c$ is the generator of $\mathbf{T}^c(\tau)$.

\paragraph*{Lemma} If the $n_c$ eigenfunctions $\mathcal{F}=\lbrace \mathcal{F}_1,...,\mathcal{F}_{n_c} \rbrace$ of $\mathcal{Q}$, associated to eigenvalues $\theta_j \approx 0$, are $\pi$-orthonormal and $\chi = \mathcal{F} \mathcal{A}$ is a regular basis transformation of these eigenfunctions, then 
\[
\mathrm{if} \,\  \exp(\tau \, \mathcal{Q}) = \mathcal{T}(\tau) \, \Rightarrow \, \exp(\tau \, \mathbf{Q}^c) = \mathbf{T}^c(\tau) 
\]
where $\mathcal{A}$ is a linear transformation corresponding to the matrix $A$.

\textit{Proof.} 
The eigenfunctions $\mathcal{F}_j$ of $\mathcal{Q}$ are eigenfunctions also of $\mathcal{T}(\tau)$ with eigenvalues $\lambda_j(\tau)=\exp(\tau \, \theta_j)$ (semigroup property). 
\begin{equation}
\begin{aligned}
\mathcal{T}(t) \mathcal{F}_j = &  \lambda_j(\tau) \mathcal{F}_j = \exp(\tau \, \theta_j) \mathcal{F}_j \\
\mathcal{Q} \, \mathcal{F}_j = & \theta_j \, \mathcal{F}_j
\end{aligned}
\label{eq:p1}
\end{equation}
By replacing $\chi = \mathcal{A}^\top \mathcal{F}^\top$ and  $\chi= \mathcal{F} \mathcal{A}$ in eq.\ref{conf_trans}, then we have
\begin{equation}
\begin{aligned}
\mathbf{Q}^c = & \frac{\langle  \chi, \mathcal{Q} \, \chi  \rangle_\pi }{\langle \chi, \chi \rangle_\pi} \\
= &\frac{\mathcal{A}^\top \, \langle \mathcal{F}^\top ,\, \mathcal{Q}  \mathcal{F} \rangle_\pi \, \mathcal{A}}{\mathcal{A}^\top \, \langle \mathcal{F}^\top, \, \mathcal{F} \rangle_\pi \, \mathcal{A}} \\
= &\frac{\mathcal{A}^\top \, \langle \mathcal{F}^\top ,\, \Theta \mathcal{F} \rangle_\pi \, \mathcal{A}}{\mathcal{A}^\top \, \mathcal{A}} \\
= & \mathcal{A}^\top \, \langle \mathcal{F}^\top,\,  \mathcal{F} \rangle_\pi \,  \Theta \mathcal{A} \\
= & \mathcal{A}^\top \,  \Theta \, \mathcal{A} \\
\end{aligned}
\label{eq:p2}
\end{equation}
In the same way, it is possible to prove that $\mathbf{T}^c(\tau) = \mathcal{A}^\top \,  \Lambda (\tau)\, \mathcal{A}$. The matrices $\Theta$ and $\Lambda(\tau)$ denote respectively the diagonal matrices $n_c \times n_c$ of the eigenvalues $\theta_j$ and $\lambda(\tau)_j$. By exploiting the relation in eq. \eqref{eq:p1}, it yields
\begin{equation}
  \exp(\tau \, \mathcal{A}^\top \, \Theta \, \mathcal{A})  = \exp(\tau \, \mathbf{Q}^c)  = \mathbf{T}^c(\tau) =  \mathcal{A}^\top \,  \Lambda(\tau) \, \mathcal{A} 
\label{eq:lemma}
\end{equation}
Then $\mathbf{Q}^c$ is an infinitesimal generator of $\mathbf{T}^c(\tau)$.
\hfill$\square$

The above lemma also provides a method to estimate the transition rate matrix of the conformations. The linear transformation $\mathcal{A}$ is unknown, however, by  Robust Perron Cluster Analysis, it is possible to estimate the optimal matrix $\mathbf{A}$ starting from the rate matrix $Q$ obtained by Square Root Approximation.

\section{Flux estimation}
In the theory section \ref{sec:sqra}, we derived the approximation \eqref{eq:sqra_final} of the transition rate between two neighboring states $i$ and $j$, i.e. $Q_{ij}''' = \sqrt{\frac{\pi_j}{\pi_i}} \hat \Phi$. The quantity $\hat \Phi$ denotes the flux trough the intersecting surface, a quantity that does not depend on the potential energy function or on the cells $\Omega_i$ and $\Omega_j$. It is hard to estimate this quantity  numerically, because of the curse of dimensionality.
In this article, we propose a method that exploits the relation $\exp(\tau \, \mathbf{Q}^c)  = \mathbf{T}^c(\tau)$ to estimate the flux.

Given the time-discrete trajectory $x_t$ of a dynamical system governed by a Hamiltonian function $\mathcal{H}(x)$ that well samples the state space according to the Boltzmann distribution $\pi(x) = \exp \left( - \beta \mathcal{H}(x) \right)/Z$. 
From this sampling, we discretize the state space with a Voronoi tessellation and we construct the matrix 
\begin{equation}
\widetilde Q_{ij} = \sqrt{\frac{\pi_j}{\pi_i}} = \frac{Q_{ij}'''}{\hat \Phi}
\label{eq:sqra_noflux}
\end{equation} 
that differs from the rate matrix in equation \eqref{eq:sqra_final} because we are neglecting the constant flux. We remark that $i$ and $j$ are the indices of neighbor cells.
The same trajectory can  also be used to construct a Markov State Model\cite{Prinz2011}, a method to estimate the transition probability matrix $\mathbf{T}(\tau)$. 

By PCCA+ we reduce both of the matrices $\widetilde{\mathbf{Q}}$ and $\mathbf{T}(\tau)$, to the transition matrices $\widetilde{\mathbf{Q}}^c$ and $\mathbf{T}^c(\tau)$ between the conformations. Thus we obtain
\begin{equation}
T_{ij}^c(\tau) = \exp \left( \tau \, Q_{ij}'''^c\right) = \exp \left( \tau \, \hat \Phi \, \widetilde Q_{ij}^c \right)
\end{equation}
If $\widetilde{\theta}_i$ are the eigenvalues of $\widetilde{\mathbf{Q}}^c$ and $\lambda_i(\tau)$ are the eigenvalues of $\mathbf{T}^c(\tau)$, the flux is given as
\begin{equation}
\hat \Phi = \frac{\log \lambda_i(\tau)}{\tau \, \widetilde{\theta}_i} \quad \forall i
\label{eq:flux}
\end{equation}
The flux should not depend on the choice of the eigenvalues. Nevertheless, in section \ref{sec:results} we will see, that depending on the quality of the Voronoi discretization, the fluxes obtained from different eigenvalues converge or not to the same value.
This fact is useful to evaluate the quality of the rate matrix.

We remark that the constant $\hat \Phi$ found in eq. \eqref{eq:flux}, from a mathematical point of view is the same constant $\hat \Phi$ that appears in eq. \eqref{eq:sqra_final}. On the other hand this quantity has a different physical interpretation. 
Indeed, the conformation matrix $\widetilde{\mathbf{Q}}^c$ describes transitions between metastable states, while the matrix $\widetilde{\mathbf{Q}}$ describes transitions between neighboring states.

\subsection{Two dimensional system}
\begin{figure*}[h!]
  \begin{center}
  \includegraphics[scale=1]{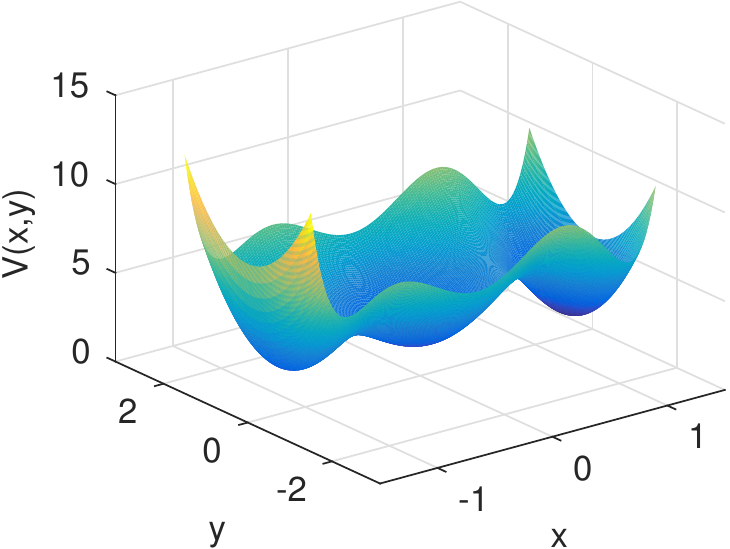}
    \caption{Diffusion process 2d. Potential energy function $V(x,y)$.}
    \label{fig:pot_en}
  \end{center}
\end{figure*}

We consider a two-dimensional convection-diffusion process of the stochastic differential equation:
\begin{equation}
\begin{cases}
\mathrm{d} x_t=-\nabla_x V\left(x_t, y_t\right) + \sigma \mathrm{d} B_t^x \\
\mathrm{d} y_t=-\nabla_y V\left(x_t, y_t\right) + \sigma \mathrm{d} B_t^y
\end{cases}
\end{equation}
where $B_t^i$ denotes a standard Brownian motion in the direction $i=x,y$, $\sigma=1,1.5,2, 2.5$ is the volatility and $V\left(x_t, y_t\right)$ is a two-dimensional potential energy surface given by the function
\begin{equation}
V(x,y) = 4 \left(x^3 - \frac{3}{2}x \right)^2 - x^3 + x + 2 + 2y^2
\label{eq:pot_function}
\end{equation}
This function describes a two-dimensional triple-well potential along the direction $x$ as reported in fig. \eqref{fig:pot_en} and guarantees the presence of three metastable states.
The system has been solved by using the Euler-Maruyama scheme:
\begin{equation}
\begin{cases}
x_{n+1}=x_n -\nabla_x V\left(x_n, y_n \right)\Delta t + \sigma  \eta^x \sqrt{\Delta t}\\
y_{n+1}=y_n -\nabla_y V\left(x_n, y_n \right)\Delta t + \sigma  \eta^y \sqrt{\Delta t}\\
\end{cases}
\end{equation}
where $\Delta t=0.001$ is the integration time-step and $\eta^i$ are i.i.d random variables drawn from a standard Gaussian distribution. We have produced trajectories of $4 \times 10^7$ time-steps.

\begin{figure*}[h!]
  \begin{center}
  \includegraphics[scale=1]{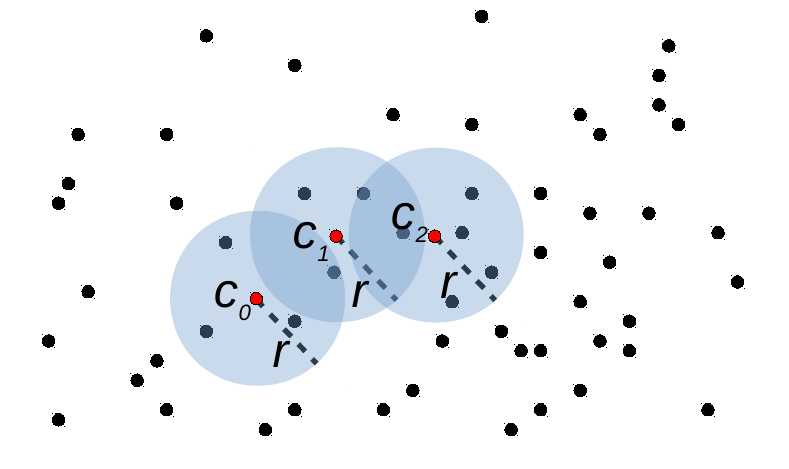}
    \caption{Explanation of the algorithm to select the states of the Voronoi tessellation. The black points represent the states of a trajectory, to select those are becoming centers of the Voronoi cells, we have set a minimum distance $r$. The first center $c_0$ is randomly chosen, the next center $c_1$ is the nearest out of a radius $r$. The next centers are found iteratively.  }
    \label{fig:parameter_r}
  \end{center}
\end{figure*}

The Voronoi tessellation has been constructed such that the cells are of approximately the same size. 
We chose a random initial point $c_0$ of the trajectory as first center of a cell, and fixed a minimum distance $r$. To select the next center $c_1$, we have picked the point of the trajectory nearest to $c_0$ outside the radius $r$.
Then we iterated this procedure to search the next centers. Fig. \ref{fig:parameter_r} describes the algorithm.
Note that one can speed up the search, by removing all points from the list that are within a radius $r$ from the already found centers. 

This algorithm guarantees the cells are homogeneously distributed and have approximately the same area. The final number of cells depends on the temperature of the system. A simulation at high temperature samples a bigger state space than a simulation at low temperature, thus also the number of Voronoi cells, whose volume depend on the variable $r$, increases.
We have used $r = 0.1, 0.15, 0.2$ that yield between $400$ and $2700$ cells depending on the volatility (temperature) of the system. 

The set of the points $c_i$ is used to construct the Voronoi tessellation. Actually this operation could be avoided, because we need to know only the adjacency matrix that identifies the neighboring cells. 
Thinking the Voronoi diagram in term of convex polyhedra\cite{Komei2004,Lie2013} permits to write a Linear Program to estimate the adjacency matrix, improving the efficiency compared to usual algorithms.

The rate matrix $\widetilde{\mathbf{Q}}$ has been constructed according to the equation \eqref{eq:sqra_noflux} with $\beta = 2/\sigma^2$.
The MSM has been constructed on the same tessellation choosing a lag time range from 100 to 1000 time steps.
The PCCA+ analysis has been realized by using $n_c=3$ (the number of mestable states) as input parameter.

To study the dependence of the flux on the potential energy, we have perturbed the potential energy function \eqref{eq:pot_function} by adding the function
\begin{equation}
U(\kappa,x) = \kappa x
\label{eq:perturbation}
\end{equation}
where $\kappa$ is a parameter that tunes the strength of the perturbation.

\subsection{Alanine dipeptide}
We performed all-atom MD simulations of acetyl-alanine-methylamide (Ac-A-NHMe, alanine dipeptide) in explicit water. All simulations were carried out with the GROMACS 5.0.2 simulation package\citep{Spoel2005} with the force field AMBER ff-99SB-ildn \citep{Larsen2010} and the TIP3P water model \citep{Jorgensen1983}.
We have performed simulations in a NVT ensemble, at temperature of 900 K and at temperture of 300 K. The length of each simulation was 600 ns and we printed out the positions every \textsf{nstxout}=500 time steps, corresponding to 1 ps. A velocity rescale thermostat has been applied to control the temperature and a leapfrog integrator has been used to integrate the equation of the motion.
To discretize the state space with a Voronoi tesselletion, we used the same algorithm described for the two-dimensional system. The same discretization has been used also to construct the MSM.

\section{Results and discussion}\label{sec:results}
\subsection{Two dimensional system}
As first application, we consider a two dimensional diffusion process. This example is important to highlight the main properties of the rate matrix estimated by square root approximation. 
The potential energy function (eq. \eqref{eq:pot_function}, fig. \ref{fig:pot_en}) has three minima respectively at (-1.12,0), (0.05, 0) and (1.29,0), separated by three barriers, whose highest points are approximatively at (-0.83, 0) and (0.61,0).

\begin{figure*}[h!]
  \begin{center}
  \includegraphics[scale=1]{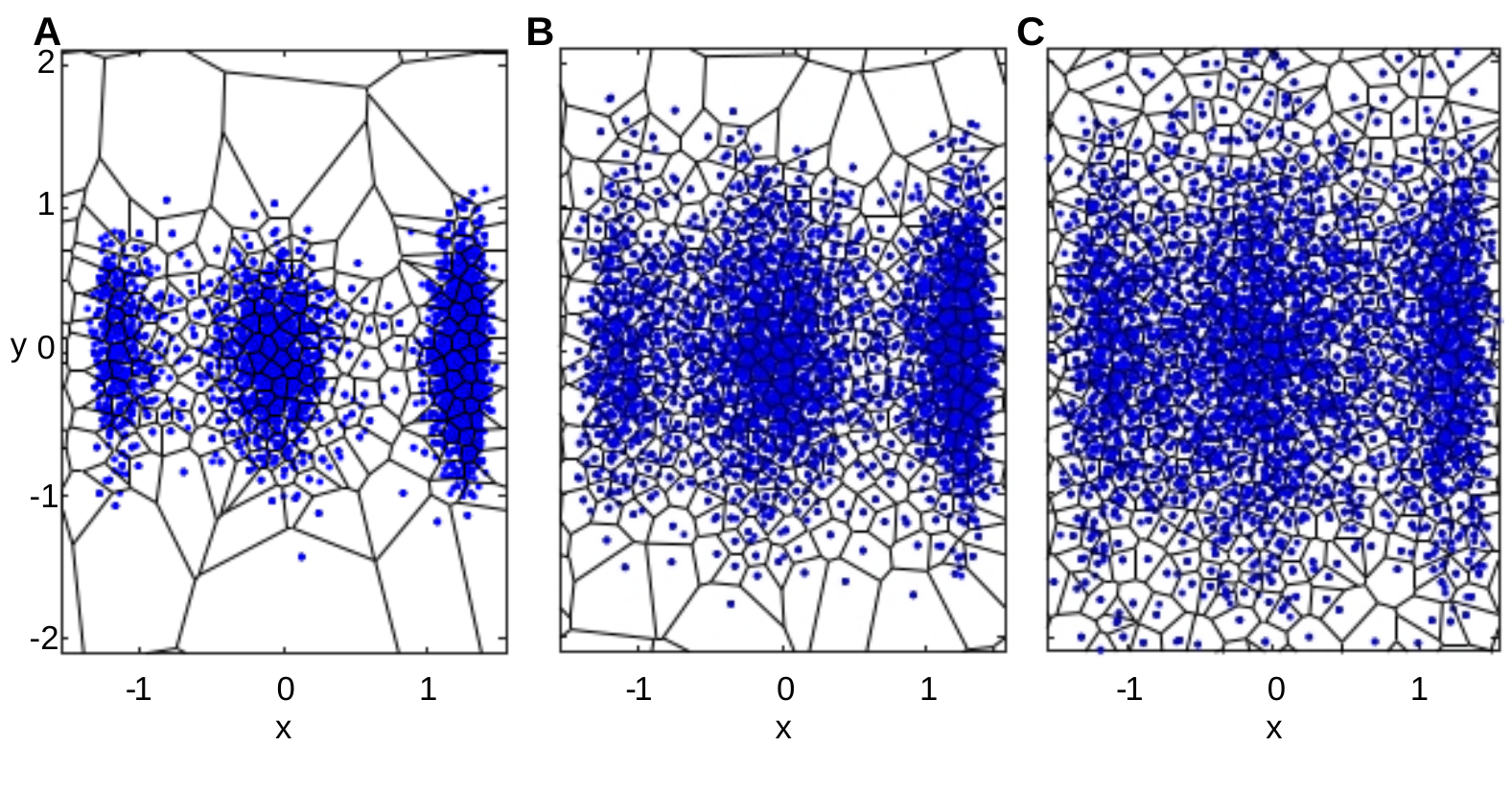}
    \caption{Two-dimensional system. Trajectories $(x_n,y_n)$ generated with $\sigma=1.0$ (A), $\sigma=1.5$ (B) and $\sigma=2.0$ (C). The Voronoi tessellation are made respectively by 251, 452 and 605 cells. To realize the pictures, we picked $4\times 10^4$ points, every 1000 timesteps, from long trajectories of $4\times 10^7$ points.  }
    \label{fig:voronoi}
  \end{center}
\end{figure*}

Fig. \ref{fig:voronoi} shows three different trajectories generated respectively with volatility $\sigma=1.0$, $\sigma=1.5$ and $\sigma=2.0$ together with the Voronoi discretization of the space, keeping constant the parameter $r$, i.e. the minimum distance between the centers of neighboring cells. Note that this parameter is also related to the number of cells and to their volume. In particular, if the state space is uniformly sampled, the minimum distance between neighboring centers turns out to be approximately the average distance between neighboring centers. This guarantees also that the cells have on average the same volume.
By increasing the volatility, that represents the temperature of the environment, the trajectory covers a larger area, thus the Voronoi tessellation results to have more cells, distributed more homogeneously and with almost the same volume. As we will see, this fact has big impact on the construction of the rate matrix.

\begin{figure*}[h!]
  \begin{center}
  \includegraphics[scale=1]{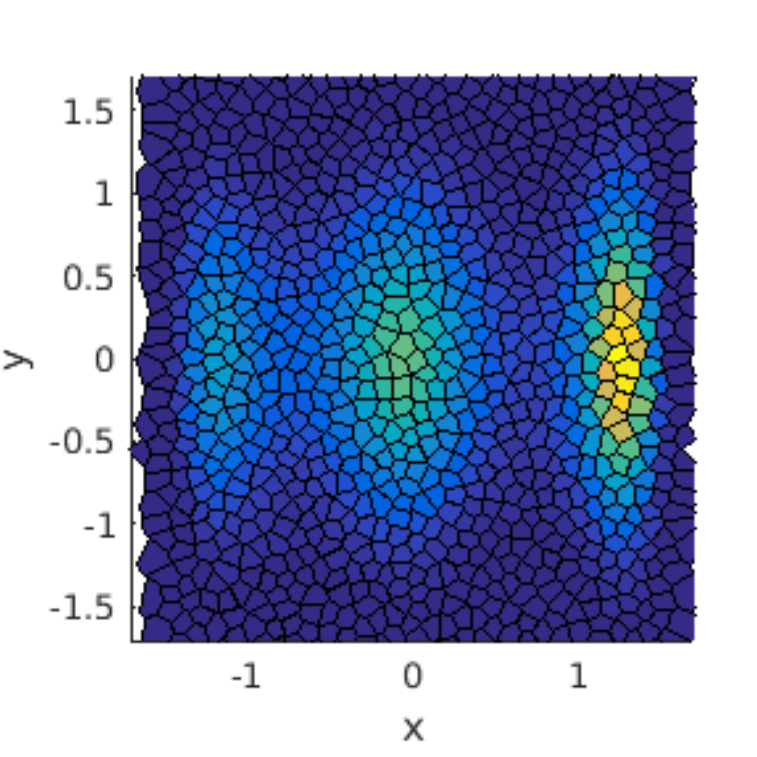}
    \caption{Two-dimensional system. Boltzmann density of the Voronoi cells (\textsf{ncells=}1258). The gradient color, from blue to yellow denotes the density of the cells. The trajectory used to construct the rate matrix has been generated with $\sigma=1.5$ and $r=0.1$.}
    \label{fig:b_weights_2d}
  \end{center}
\end{figure*}

We start discussing the results of a simulation realized with volatility $\sigma=1.5$, after having discretized the space in 1258 Voronoi cells, using a minimum distance between centers of neighbor cells $r=0.1$ and taking into account periodic boundary condition.
Fig. \ref{fig:b_weights_2d} shows the Boltzmann weights of the Voronoi cells, estimated as average over all the points falling in each cell:
\begin{equation}
\pi_i = \frac{1}{\textsf{nsteps}}\sum_{n=1}^{\textsf{nsteps}}\mathbf{1}_{\Omega_i}(x_n,y_n) \, \exp \left( - \beta V(x_n,y_n) \right) \quad \forall \, \textrm{cell}\, \Omega_i
\end{equation}
where $\mathbf{1}_{\Omega_i}(x_n,y_n)$ is the indicator function and \textsf{nsteps} is the length of the trajectory. The picture of the Boltzmann weights represents the distribution of the most visited states and it is specular to the potential energy function. Thus, we observe high values in correspondence of the potential minima, and low values in correspondence of the barriers.
The Boltzmann weights have been used to estimate the matrix \eqref{eq:sqra_noflux} $\widetilde{\mathbf{Q}}$ that, by definition is proportional to the rate matrix $\mathbf{Q}'''$ in eq. \eqref{eq:rate_matrix_sqra}, although we still do not know the value of the proportionality constant $\hat \Phi$.

\begin{figure*}[h!]
  \begin{center}
  \includegraphics[scale=1]{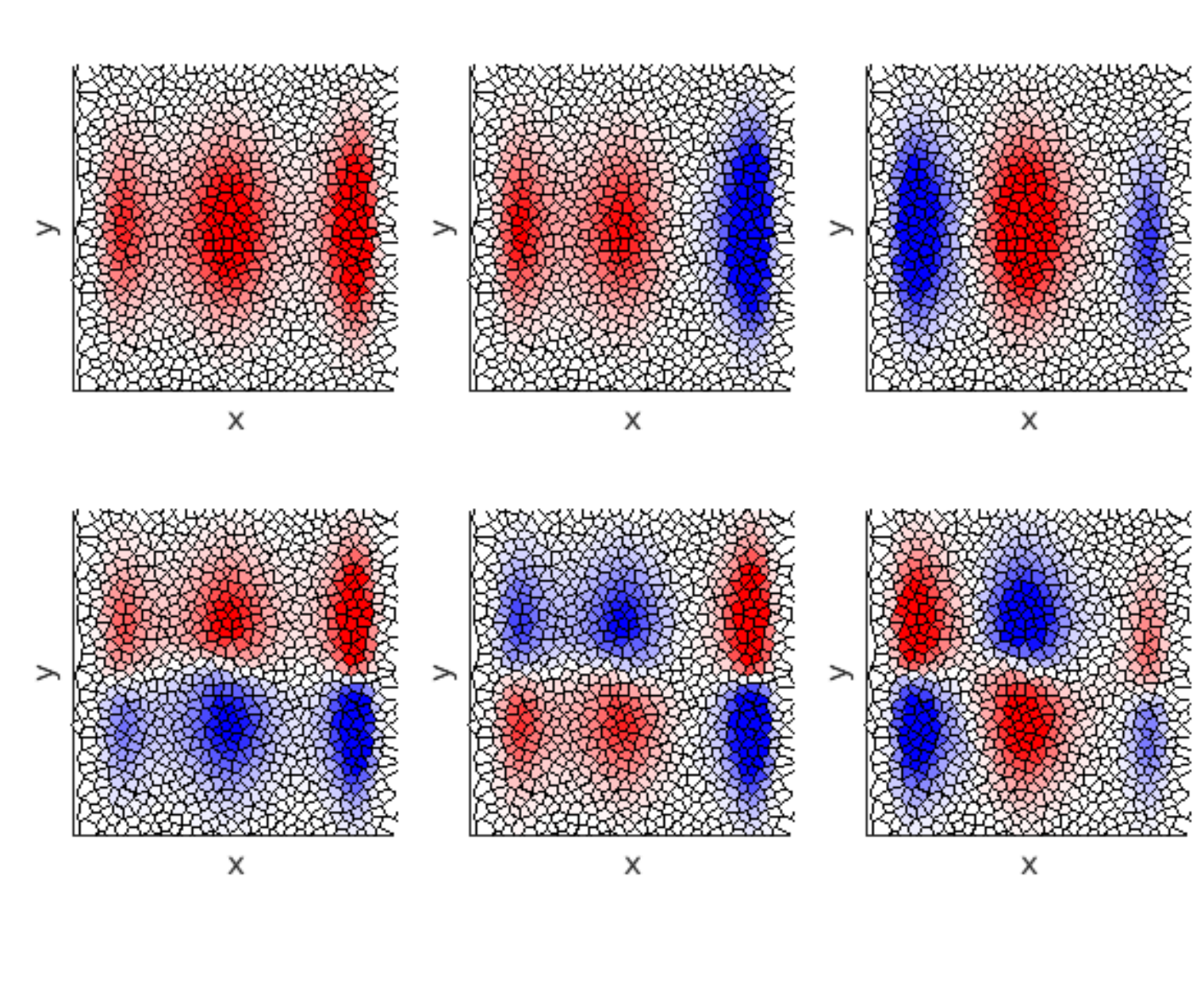}
    \caption{Two-dimensional system. First six eigenvectors of the rate matrix $\widetilde{\mathbf{Q}}$. The red color denotes positive entries, the blue color denotes negative entries. The trajectory used to construct the rate matrix has been generated with $\sigma=1.5$ and $r=0.1$.}
    \label{fig:evecs_2d}
  \end{center}
\end{figure*}

This matrix, even if it does not contain the correct rates, can be used for a qualitative analysis of the dynamics of the system.
Fig. \ref{fig:evecs_2d} shows the first six left eigenvectors of the matrix. The first eigenvector has only positive entries and is proportional to the Boltzmann weights in fig. \ref{fig:b_weights_2d}.
The second eigenvector contains also negative values and represents the slowest transition of the dynamics that happens between the third minimum (the deepest minimum) and one of the two left minima.
The third eigenvector represents the second slowest transition between the central and external minima. The next eigenvectors describe the faster processes that can happen even between states belonging to the same metastable state.

The eigenvalues of the matrix contain information about the implied timescales, i.e. the average times at which the transitions occur. The first eigenvalue is zero, according to the Perron-Frobenius theorem, while the others are negative.
It is known\cite{Prinz2011} that the implied timescales of a Markovian process, described by a transition probability matrix $\mathbf{T}(\tau)$ are obtained as
\begin{equation}
t_i = - \frac{\tau}{\log \lambda_i(\tau)}
\label{eq:its_lambda}
\end{equation}

where $\lambda_i(\tau)$ is th $i$th eigenvalue of the matrix $\mathbf{T}(\tau)$. Thus, exploiting the semigroup property \eqref{eq:p1}, a simple calculation
shows that the implied timescales can be obtained as 
\begin{equation}
t_i = - \frac{1}{\hat \Phi \widetilde{\theta_i}}
\label{eq:its_theta}
\end{equation}
where $\widetilde{\theta_i}$ are the eigenvalues of the rate matrix $\widetilde{\mathbf{Q}}$.
The advantage of the relation \eqref{eq:its_theta} respect to \eqref{eq:its_lambda} is that one does not need anymore to check the convergence of the implied timescales as in Markov State Model construction \cite{Prinz2011}, because eq. \eqref{eq:its_theta}  does not depend on the lag-time $\tau$.
On the other hand we do not know the value of $\hat \Phi$ yet and thus we can only relatively compare the transitions one to another.
For example, the first slow transition happens at a timescale of $t_1 = 70.53 \, \hat\Phi^{-1}$ time steps, while the second transition at $t_2 = 19.14 \hat\Phi^{-1}$ time steps, then the second transition is $3.68$ times faster than the first one. Table \ref{table:table_its} contains the first five implied timescales, while the ratios between the implied timescales are in table \ref{table:table_ratios}.

To estimate the value of the flux, we have constructed the transition probability matrix by Markov State Model, then we have reduced both the matrices $\widetilde{\mathbf{Q}}$ and $\mathbf{T}(\tau)$ by PCCA+, to the conformation matrices $\widetilde{\mathbf{Q}}^c$ and $\mathbf{T}^c(\tau)$.
Afterwards, we have estimated the constant $\hat \Phi$ by studying the eigenvalues, according to equation \eqref{eq:flux}.

Because the system has three metastable states, the matrices $\widetilde{\mathbf{Q}}^c$ and $\mathbf{T}^c(\tau)$ are $3\times 3$, thus  the constant $\hat \Phi$ can be estimated comparing the eigenvalues $\widetilde{\theta}_i$ and $\lambda_i(\tau)$ with $i=2,3$. 
Tables \ref{table:table1}, \ref{table:table2}, \ref{table:table3} and \ref{table:table4} contain the values of $\hat \Phi$ for several simulations with different initial conditions. In each table, $\hat \Phi_2$ denotes the flux estimated comparing the second eigenvalues, $\hat \Phi_3$ denotes the flux estimated comparing the third eigenvalues, $\bar \Phi$ is the average flux between $\hat \Phi_2$ and $\hat \Phi_3$, "std" is the standard deviation and "rel. err." is the relative error.

Table \ref{table:table1} reports the flux as function of the volatility $\sigma$, i.e. the temperature of the environment. We observe that the average flux increases linearly. 
Indeed at high temperature, the dynamics is faster and a single cell is visited for a short amount of time.
Furthermore, the trajectory covers also those zones difficult to sample due to a high gradient of the potential energy function. As consequence, the number of Voronoi cells grows up (keeping constant the parameter $r$) and we observe a reduction of the relative error.

In table \ref{table:table2} we observe how the flux depends on the choice of the parameter $r$, i.e. on the minimum distance between the centers of neighboring cells. By reducing $r$ (and consequently the volume of the cells), the average flux increases. Indeed by reducing the size of the cells, a cell is visited for a minor amount of time and the number of transitions trough the intersecting surfaces grows up. We observe also an important reduction of the relative error that denotes the convergence of the rate matrix.

Finally, we have studied the dependence of the flux on the external perturbation of the potential energy function described in eq. \eqref{eq:perturbation} with $\kappa=0, \,0.5, \, 1$. The effect of such perturbation is to tilt the potential energy along the axis $x$, and consequently it redistributes the Boltzmann weights. The experiment has been repeated for $\sigma=1.5$ and $\sigma=2.0$ and the results are collected respectively in table \ref{table:table3} and \ref{table:table4}. In both the cases the perturbation does not affect the average flux. This result confirms the initial assumption in section \ref{sec:sqra}, that the flux does not depend on the potential energy function.
\subsection{Alanine dipeptide}
We have repeated a similar analysis for alanine dipeptide (Ac-A-NHMe) in explicit water.
The main difference with respect to the 2d system is that the simulation has been carried out on all the degrees of freedom (all atomic simulation), but we have constructed the rate matrix (and the MSM) only on two relevant coordinates, the backbone dihedral angles $\phi$ and $\psi$, that well capture the main dynamical properties of the system. This approximation requires to fix the proposed approach to obtain the rate matrix in a low dimension space.

First of all, to get the correct value of the Boltzmann weights one should use the potential energy function $V(\mathbf{r})$ (force field) in full dimension. However, this would return an unreadable result due to the high dimensionality of the system, thus we have replaced the true Boltzmann weights in full dimension, with a normalized histogram on the dihedral angles $\phi$ and $\psi$.
This assumption is correct if the relevant coordinates well represent the stationary distribution of the system. In such case, we have
\begin{equation}
\pi_i = \mathbf{1}_{\Omega_i}(\mathbf{r}) \, \frac{\exp \left( - \beta V(\mathbf{r}) \right)}{Z} \propto h_i = \frac{\sum_{n=1}^\textsf{nsteps}\mathbf{1}_{\Omega_i}(\phi_n,\psi_n)}{\textsf{nsteps}} \quad \forall \, \mathrm{cell} \, \Omega_i
\end{equation}
where $\pi_i$ is the true Boltzmann weight of the cell $\Omega_i$, $Z$ is the unknown partition function, $\lbrace \phi_n, \psi_n \rbrace$ is the state of the trajectory at time step $n$ and \textsf{nsteps} is the length of the trajectory.
The histogram can be used also to estimate the free energy profile $F(\phi, \psi)$, defined up to a negligible constant and to rescale the Boltzmann weights according to a new temperature. 
\begin{equation}
\begin{aligned}
h_i = & \frac{\exp \left( - \beta \, F_i \right)}{N} \quad \forall \, \mathrm{cell} \, \Omega_i \\ 
F_i = & - \frac{1}{\beta} \log h_i + \textrm{const.}
\end{aligned}
\end{equation}
where $F_i$ is the free energy of the cell $\Omega_i$ in the space $\lbrace \phi_n, \psi_n \rbrace$ and $N$ is a normalization constant. 
The new Boltzmann weights, for a different temperature $\beta' = 1/(k_B T')$ would be
\begin{equation}
h_i' = \frac{\exp \left( - \beta' \, F_i \right)}{N} \quad \forall \, \mathrm{cell} \, \Omega_i 
\label{eq:rate_rew}
\end{equation}

\begin{figure*}[h!]
  \begin{center}
  \includegraphics[scale=1]{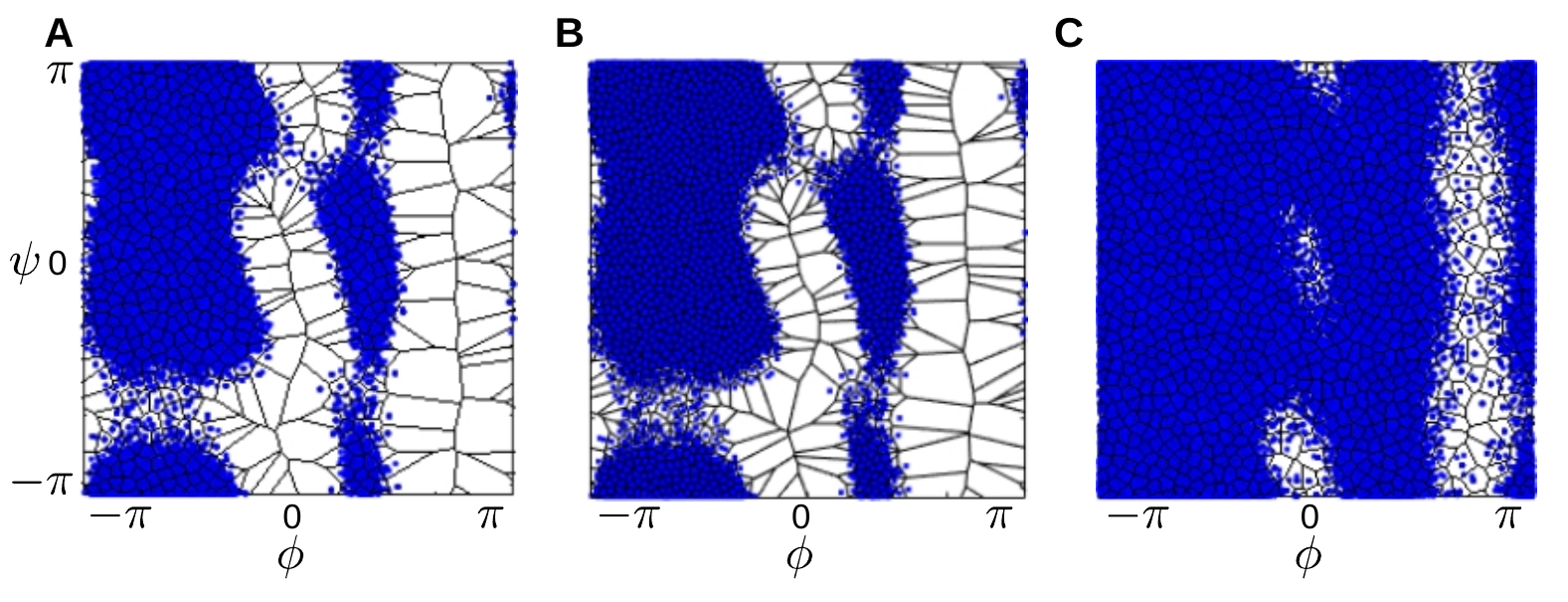}
    \caption{Alanine dipeptide. Trajectories $(\phi_n,\psi_n)$ generated at temperature 300 K (A) and (B), at 900 K (C). The Voronoi tessellation are made respectively by 566 (A), 1433 (B) and 1005 cells, using $r=0.17$ (A and C), $r=0.1$ (B).  }
    \label{fig:voronoi_ala}
  \end{center}
\end{figure*}

Let's now discuss the results. Fig.~\ref{fig:voronoi_ala} shows respectively two trajectories generated at temperature 300 K (A and B) and 900 K (C), projected on the backbone dihedral angles $\phi$ and $\psi$. At temperature 300 K, the trajectory does not cover the full space and the Voronoi tessellation presents cells with different size. Reducing the parameter $r$, does not improve the discretization. Even though the number of cells grows (and the size reduces), there are still cells with very large volume. The only way to improve the discretization, i.e. to have small cells with almost the same volume, is to increase the temperature to promote a better sampling of the state space. Indeed, the trajectory at temperature 900 K covers a larger subspace of the state space, that turns out in a better Voronoi discretization of the space.

\begin{figure*}[h!]
  \begin{center}
  \includegraphics[scale=1]{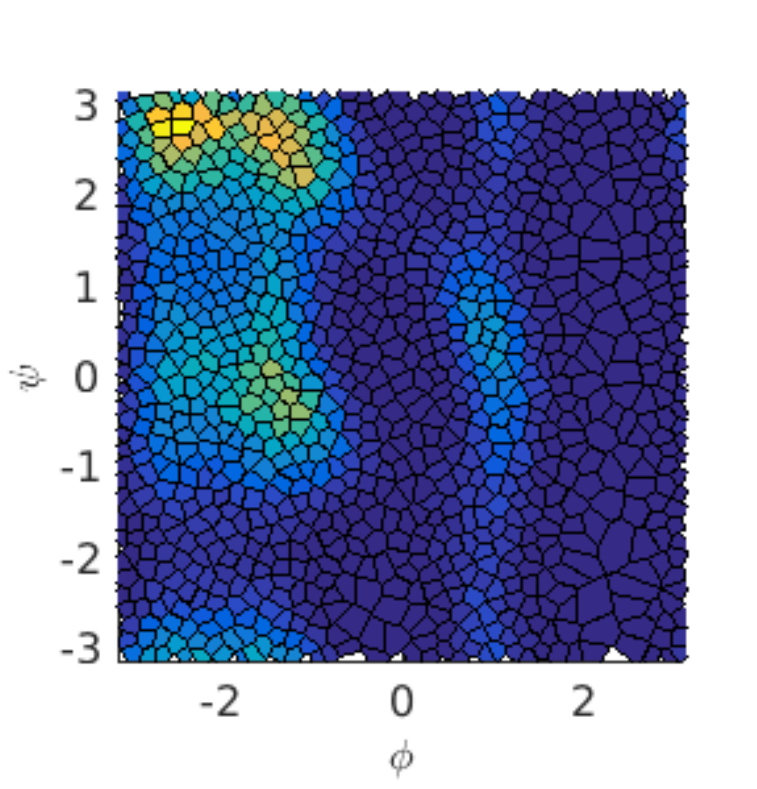}
    \caption{Alanine dipeptide. Boltzmann density of the Voronoi cells (\textsf{ncells=}1005). The gradient color, from blue to yellow denotes the density of the cells. The trajectory used to construct the rate matrix has been generated at 900 K and $r=0.17$.}
    \label{fig:b_weights_ala}
  \end{center}
\end{figure*}

\begin{figure*}[h!]
  \begin{center}
  \includegraphics[scale=1]{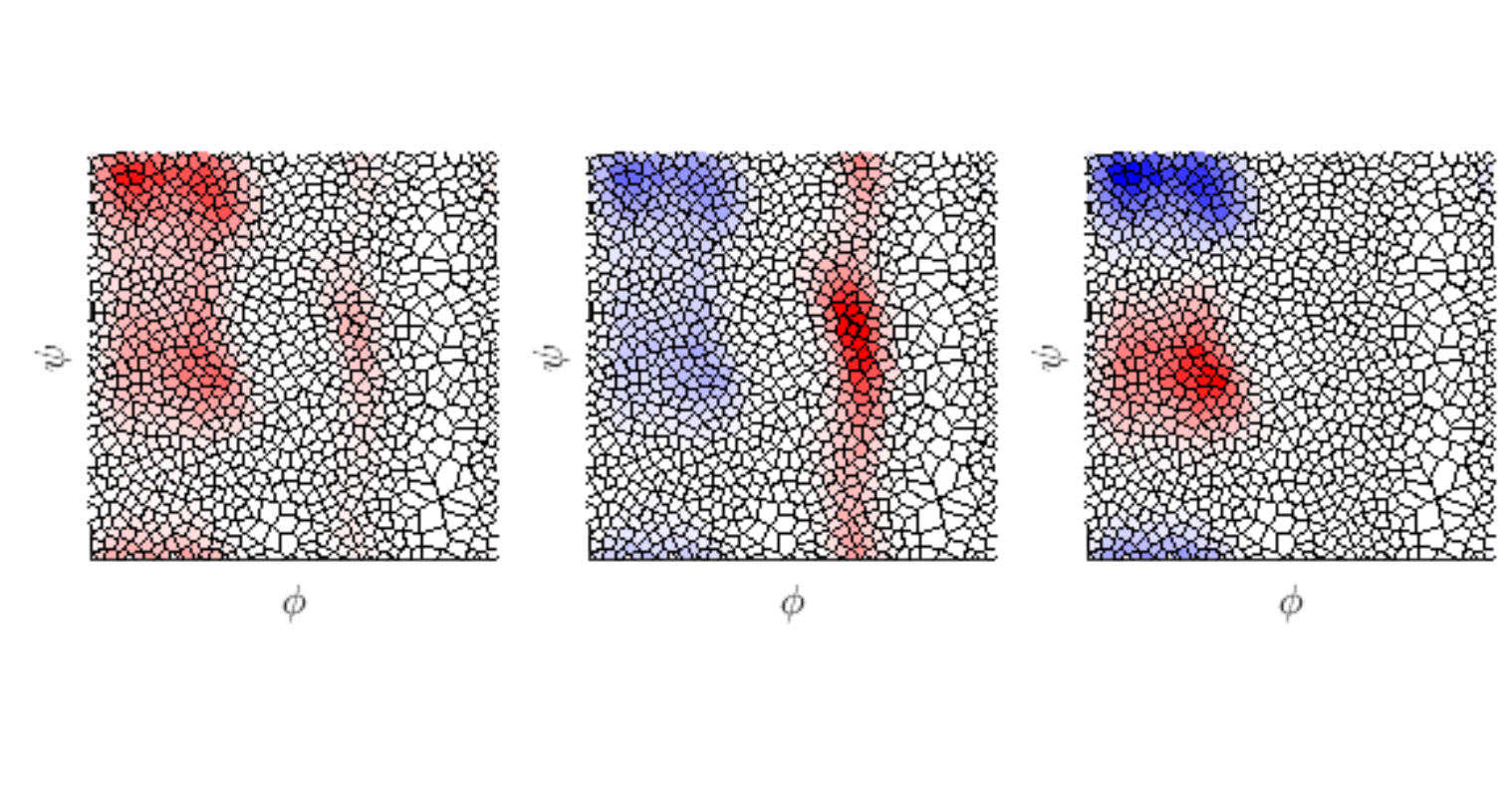}
    \caption{Alanine dipeptide. First three eigenvectors of the rate matrix $\widetilde{\mathbf{Q}}$. The red color denotes positive entries, the blue color denotes negative entries. The trajectory used to construct the rate matrix has been generated at 900 K and $r=0.17$.}
    \label{fig:evecs_ala}
  \end{center}
\end{figure*}

Fig.~\ref{fig:b_weights_ala} shows the Boltzmann weights at temperature 900 K, that
correspond to the typical equilibrium distribution in the Ramachandran plane. 
Fig.~\ref{fig:evecs_ala} shows the dominant left eigenvectors of the matrix $\widetilde{Q}_{ij} = \sqrt{h_j/h_i}$. The entries of the first eigenvector are only positive and proportional to the Boltzmann weights $h_i$.
The second eigenvector represents a torsion around the $\phi$ angle and corresponds to a kinetic exchange between the L$_{\alpha}$-minimum ($\phi>0$) and the $\alpha$-helix and $\beta$-sheet minima ($\phi < 0$). The associated transition, according to eq. \eqref{eq:its_theta}, is $146 \hat \Phi^{-1}$ ps. 
The third eigenvector represents a transition $\beta$-sheet $ \longleftrightarrow \alpha$-helical conformation, i.e. a torsion around $\psi$, and is associated to a timescale of $40 \hat \Phi^{-1}$ ps.
The eigenvectors match very well with the eigenvectors that one would obtain from a MSM\cite{Vitalini2015}, however we do not know the value of the constant $\hat \Phi$, thus the timescales need to be corrected.

As in the two-dimensional case, we have constructed a MSM, then we have produced the conformation matrices assuming the existence of three metastable states ($\beta$-sheet, L$\alpha$-helix and R$\alpha$-helix) and we have compared the eigenvalues.

Table \ref{table:table5} collects the results for different experiments. In the first two rows of the table, we used the same trajectory at high temperature, to construct the rate matrix and the MSM. We observe that increasing the number of cells (i.e. by reducing their volume), the average flux grows up and the precision improves, thus we have that $\hat \Phi_2  \simeq  \hat \Phi_3$.
This result confirms what we already discussed for the two-dimensional diffusion process.

The third, forth and fifth rows contain the results for a trajectory generated at 300 K. In this case we observe that by increasing the number of cells causes an increase of the flux, however, the relative error is very high in all the three cases. 
As we can see in fig.~\ref{fig:voronoi_ala}.A, alanine dipeptide presents 3 high barriers (approximately at $\phi \approx 0$, $\phi \approx 2/3\pi$ and $\psi \approx -4/5\pi$) that are not well sampled at low temperature. In particular we observe very large cells covering the zones of the Ramachandran plane (corresponding to high barriers) not visited by the trajectory. Reducing the parameter $r$ (and then the size of the cells), does not improve the result (the relative error is still very high), because that parameter does not affect the size of those cells in not visited regions. This results in a violation of the initial condition and in a bad rate matrix.

The last two rows of the table show the results obtained comparing the eigenvalues of a rate matrix built on the trajectory at temperature $T=900$ K, with eigenvalues of the transition probability matrix, built on the trajectory at temperature $T=300$ K. We remark that we rescaled the rate matrix to $T=300$ K, according to equation \eqref{eq:rate_rew}.
Also in this case, by increasing the number of cells, we can observe a significant increase of the flux. The relative error is approximately halved with respect to the third, forth and fifth row.
We deduce that this improvement is due to the better rate matrix, constructed with a trajectory produced at $T=900$ K.
On the other hand, the error is still high, because the transition probability matrix has been constructed between states not well sampled (i.e. from the trajectory at  $T=300$ K).
%


\section{Conclusion}
The paper contributes to the classical molecular simulation community in three ways. It provides an easy way to estimate the rates between metastable molecular conformations. It shows that this type of discretization converges to a Fokker-Planck-operator. Finally, it shows that there is an easy mathematical relation between the discretized generator of the molecular process and the potential energy landscape.   

More detailed: Since many years the concept of transfer operators, well known in thermodynamics and quantum mechanics, has been established inside the classical molecular simulation community. New methods, such as Markov State Models have been developed, to reduce complexity and study conformational transition networks of molecular systems.
The concept of transfer operators is strongly connected to the concept of generators. Numerical methods which apply for the transfer operator also can be used for the generator, which is simply the time-derivative of the transfer operator. 
The spatial discretization of the transfer operator is a transition probability matrix of a Markov chain. The spatial discretization of the generator is a rate matrix, which is in general hard to extract from time-discretized simulation data.
Our method simply uses the Boltzmann distribution of states for dicretizing the generator.
The first underlying assumption is that we can define a continuity equation for the time-derivative of the transfer operator. Then, exploiting the Gauss theorem, we write the rate between two neighbor states as a surface integral of the flux, weighted by the Boltzmann density of the intersecting surface.
The second assumption is a constant flux, i.e, the flux does not depend on the potential energy but on the discretization of the space.
Instead of computing the Boltzmann weight of the intersecting surface of the Voronoi cells, here, it is estimated as geometric average of the Boltzmann weight of the cells. This we denoted as square root approximation.
The quality of the square root approximation depends on the discretization of the state space. We propose to use a Voronoi tessellation. For this we proved that the rate matrix converges to the generator of the Smoluchowski equation\cite{Heida2017}. In fact, the numerical and molecular examples show that the corresponding rate matrices converge to a Fokker-Planck operator as the number of cells increases.

The constant flux trough the surfaces of the Voronoi cells cannot be computed directly. By comparing the eigenvalues of the rate matrix with the corresponding eigenvalues of the transition probability matrix estimated by MSM a good estimate for the flux can be provided. This especially accounts for an increasing number of cells.

According to the theory, a fine discretization improves the final result and reduces the relative error of the flux computation.
Furthermore, we have shown that an external perturbation of the potential energy function does not affect the result, according to the second assumption.
In both the systems studied, we have observed that the quality of the rate matrix depend on the quality of the sampling.
This is in accordance with theoretical studies\cite{Heida2017}.
High temperature simulations results are more suitable to estimate the rate matrix because of the better mixing properties. A high temperature (higher entropy) also results in more homogeneous Voronoi cells.
%

\begin{acknowledgments}
This research has been funded by Deutsche Forschungsgemeinschaft
(DFG) through grant CRC 1114 \emph{Scaling Cascades in Complex Systems}, Projects A05 \emph{Probing scales in equilibrated systems by optimal nonequilibrium forcing}, B05 \emph{Origin of the scaling cascades in protein dynamics} and C05 \emph{Effective models for interfaces with many scales}.
\end{acknowledgments}


\newpage

\newpage
\section*{Tables}

\begin{table}[h!]
\setlength{\tabcolsep}{12pt}
\begin{center}
\begin{tabular}{c c c}
\hline \hline
$i$ & $\theta_i$ & I.T.S. (time steps) \\ \hline            
0   &   0       &    -                 \\ 
1   &  -0.0142  & 70.5319$ \, \hat{\Phi}^{-1}$ \\ 
2   &  -0.0522  & 19.1482$ \, \hat{\Phi}^{-1}$ \\ 
3   &  -0.0738  & 13.5452$ \, \hat{\Phi}^{-1}$ \\  
4   &  -0.0871  & 11.4772$ \, \hat{\Phi}^{-1}$ \\ 
5   &  -0.1252  &  7.9900$ \, \hat{\Phi}^{-1}$ \\ 
\hline \hline
\end{tabular}
\end{center}
\caption{Two dimensional system. First five eigenvalues and implied timescales as function of the flux. The volatility was set equal to 1.5 and the minimum distance between the centers of neighbor cells was set equal to $r=0.1$.}
\label{table:table_its}
\end{table}
\begin{table}[h!]
\setlength{\tabcolsep}{12pt}
\begin{center}
\begin{tabular}{c c c c c c}
\hline \hline
\diagbox{$i$}{$j$}  & $1$ & $2$ & $3$ & $4$ & $5$ \\ \hline  
$1$                    & 1.00  & 3.68  & 5.20  & 6.14  & 8.82  \\              
$2$                    & 0.27  & 1.00  & 1.41  & 1.66  & 2.39  \\  
$3$                    & 0.19  & 0.70  & 1.00  & 1.18  & 1.69  \\ 
$4$                    & 0.16  & 0.59  & 0.84  & 1.00  & 1.43  \\ 
$5$                    & 0.11  & 0.41  & 0.58  & 0.69  & 1.00  \\
\hline \hline
\end{tabular}
\end{center}
\caption{Two dimensional system. Ratio $ITS_i / ITS_j$ between the first five implied timescales of table \ref{table:table_its}. The volatility was set equal to 1.5 and the minimum distance between the centers of neighbor cells was set equal to $r=0.1$.}
\label{table:table_ratios}
\end{table}
\begin{table}[h!]
\setlength{\tabcolsep}{12pt}
\begin{center}
\begin{tabular}{c c c c c c c}
\hline \hline
$\sigma$ & \textsf{ncells} & $\hat \Phi_2$ & $\hat \Phi_3$ & $\bar \Phi$ & std & rel. err. \\ \hline 
1.0 &   740 & 0.0080 & 0.0161 & 0.0121 & 0.0057 & 47,11\% \\  
1.5 &  1258 & 0.0461 & 0.0523 & 0.0492 & 0.0044 &  8.94\% \\  
2.0 &  1725 & 0.0913 & 0.0931 & 0.0922 & 0.0013 &  1.41\% \\ 
2.5 &  2205 & 0.1427 & 0.1372 & 0.1400 & 0.0039 &  2.79\% \\  
\hline \hline
\end{tabular}
\end{center}
\caption{Two dimensional system. Variation of the flux as function of the volatility $\sigma$. The minimum distance between the centers of neighbor cells was set equal to $r=0.1$.}
\label{table:table1}
\end{table}
\begin{table}[h!]
\setlength{\tabcolsep}{12pt}
\begin{center}
\begin{tabular}{c c c c c c c}
\hline \hline
$r$ & \textsf{ncells} & $\hat \Phi_2$ & $\hat \Phi_3$ & $\bar \Phi$ & std & rel. err. \\ \hline  
0.20 &  456 & 0.0207 & 0.0248 & 0.0227 & 0.0029 &  12.87\% \\  
0.15 &  784 & 0.0396 & 0.0425 & 0.0410 & 0.0021 &   5.12\% \\ 
0.10 & 1725 & 0.0913 & 0.0931 & 0.0922 & 0.0013 &   1.41\% \\ 
\hline \hline
\end{tabular}
\end{center}
\caption{Two dimensional system. Variation of the flux as function of the minimum distance between the centers of neighbor cells. The volatility was set equal to 2.0.}
\label{table:table2}
\end{table}
\begin{table}[h!]
\setlength{\tabcolsep}{12pt}
\begin{center}
\begin{tabular}{c c c c c c c}
 \hline \hline
$\kappa$ & \textsf{ncells} & $\hat \Phi_2$ & $\hat \Phi_3$ & $\bar \Phi$ & std & rel. err. \\ \hline  
0.0 &  1258 & 0.0461 & 0.0523 & 0.0492 & 0.0044 &   8.94\% \\  
0.5 &  1235 & 0.0459 & 0.0502 & 0.0480 & 0.0030 &   6.25\% \\  
1.0 &  1225 & 0.0449 & 0.0526 & 0.0487 & 0.0055 &  11.29\% \\  
 \hline \hline
\end{tabular}
\end{center}
\caption{Two dimensional system. Variation of the flux as function of an external perturbation. The volatility was set equal to 1.5 and $r=0.1$.}
\label{table:table3}
\end{table}
\begin{table}[h!]
\setlength{\tabcolsep}{12pt}
\begin{center}
\begin{tabular}{c c c c c c c}
\hline \hline
$\kappa$ & \textsf{ncells} & $\hat \Phi_2$ & $\hat \Phi_3$ & $\bar \Phi$ & std & rel. err. \\ \hline  
0.0 &  1725 & 0.0913 & 0.0931 & 0.0922 & 0.0013 &   1.41\% \\  
0.5 &  1722 & 0.0927 & 0.0922 & 0.0924 & 0.0003 &   0.36\% \\ 
1.0 &  1720 & 0.0904 & 0.0926 & 0.0915 & 0.0015 &   1.64\% \\ 
\hline \hline
\end{tabular}
\end{center}
\caption{Two dimensional system. Variation of the flux as function of an external perturbation. The volatility was set equal to 2.0 and $r=0.1$.}
\label{table:table4}
\end{table}
\begin{table}[h!]
\setlength{\tabcolsep}{12pt}
\begin{center}
\begin{tabular}{c c c c c c c c }
\hline \hline
$Temp.$  (K)& $r$ & \textsf{ncells} & $\hat \Phi_2$ & $\hat \Phi_3$ & $\bar \Phi$ & std & rel. err. \\ \hline
900     & 0.20 &  740 & 11.4390 & 12.9166 & 12.1778 & 1.0448 &    8.58\% \\
900     & 0.17 & 1005 & 17.0914 & 16.5862 & 16.8388 & 0.3572 &    2.12\% \\ \hline
300     & 0.17 &  566 &  0.1238 &  2.5672 &  1.3455 & 1.7277 &  128.41\% \\ 
300     & 0.14 &  792 &  0.1645 &  3.0356 &  1.6000 & 2.0301 &  126.88\% \\ 
300     & 0.10 & 1423 &  0.3090 &  6.4278 &  3.3684 & 4.3266 &  128.45\% \\ \hline 
$\theta_i(900)\sim \lambda_i(300)$ & 0.17 & 1017 &  6.6788 &  2.5062 &  4.5925 & 2.9505 &   64.24\% \\ 
$\theta_i(900)\sim \lambda_i(300)$ & 0.10 & 2660 & 15.6359 &  6.1802 & 10.9081 & 6.6862 &   61.30\% \\
\hline \hline
\end{tabular}
\end{center}
\caption{Alanine dipeptide. Variation of the flux as function of the temperature and the parameter $r$. In the first four rows, we used the eigenvalues of the rate matrix and transition probability matrix at the same temperature. In the last two rows, we used a rate matrix estimated at temperature $T=900 K$, while the transition probability matrix was estimated at temperature $T=300 K$.}
\label{table:table5}
\end{table}

\newpage
\section*{Figures}
\begin{figure*}[h!]
  \begin{center}
  \includegraphics[scale=1]{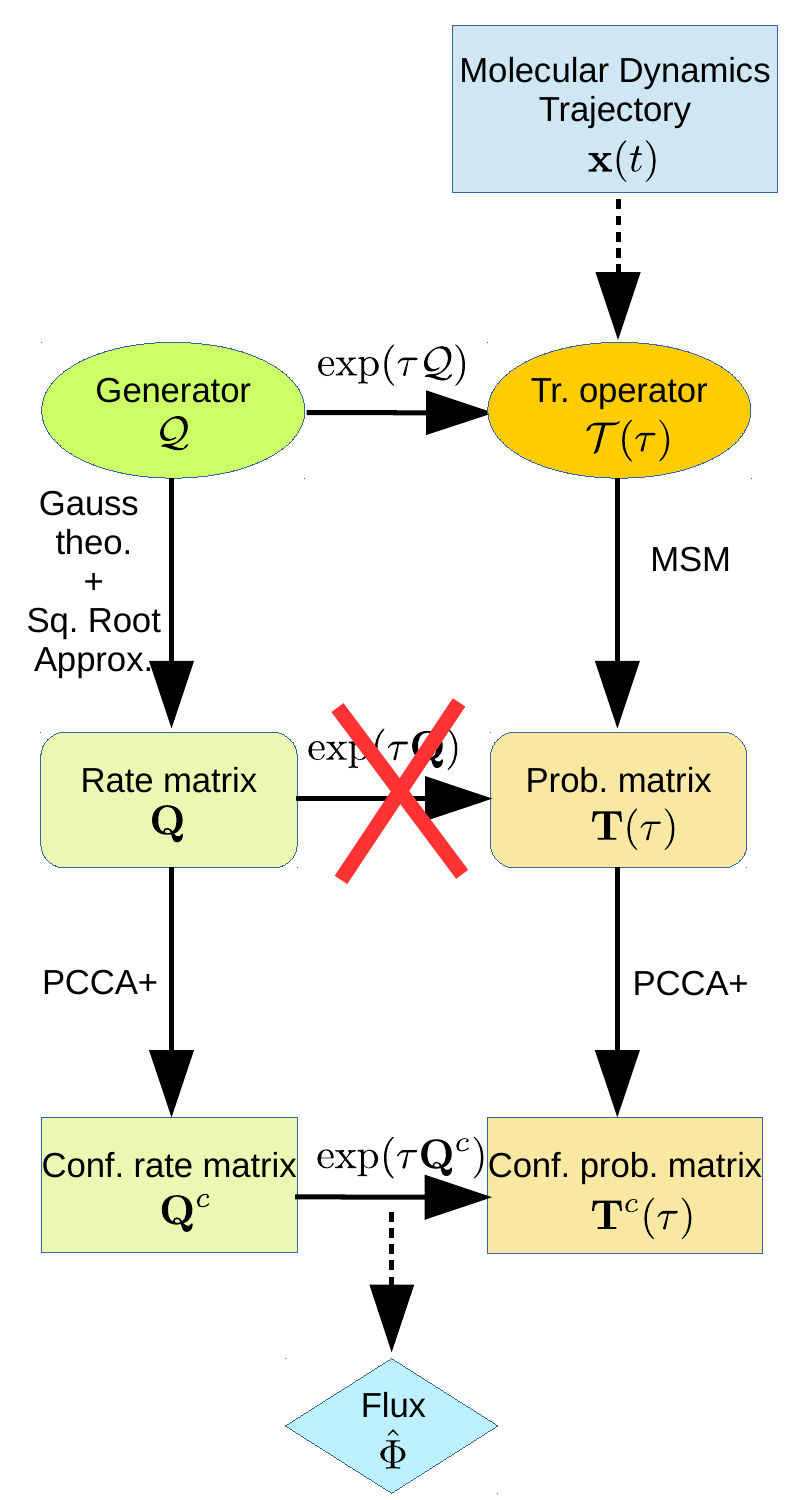}
    \caption{Scheme summarizing the theory of the transfer operator, the generator and the associated transition matrices.}
    \label{fig:scheme}
  \end{center}
\end{figure*}
\begin{figure*}[h!]
  \begin{center}
  \includegraphics[scale=1]{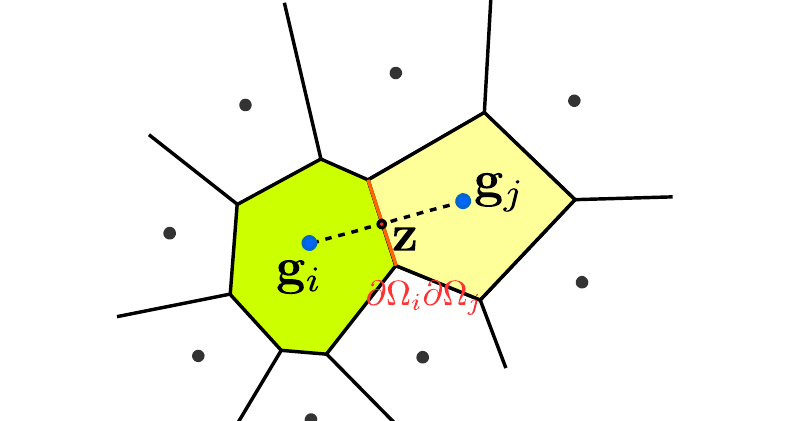}
    \caption{Two neighbouring Voronoi cells.}
    \label{fig:sqra}
  \end{center}
\end{figure*}


\end{document}